\documentclass{pasj00}

\begin{document}
\SetRunningHead{A.Tanikawa and T.Fukushige PASJ}{Effects of Hardness of Primordial Binaries}
\Received{}
\Accepted{}
\title{Effects of Hardness of Primordial Binaries on Evolution of Star
Clusters}
\author{Ataru \textsc{Tanikawa}$^{1}$ and Toshiyuki
\textsc{Fukushige}$^{2}$}
\affil{$^{1}$Center for Computational Sciences, University of Tsukuba,
1-1-1, Tennodai, Tsukuba, Ibaraki 305-8577\\
$^{2}$K\&F Computing Research Co., Chofu, Tokyo 182-0026}
\email{tanikawa@ccs.tsukuba.ac.jp}
\maketitle

\begin{abstract}
 We investigate effects of hardness of primordial binaries on whole
 evolution of star clusters by means of $N$-body simulations. Using
 newly developed code, GORILLA, we simulated eleven $N=16384$ clusters
 with primordial binaries whose binding energies are equal in each
 cluster in range of $1-300kT_0$, where $1.5kT_0$ is average stellar
 kinetic energy at the initial time. We found that, in both soft
 ($\le 3kT_0$) and hard ($\ge 300kT_0$) limits, clusters experience deep
 core collapse. In the intermediate hardness ($10-100kT_0$), the core
 collapses halt halfway due to an energy releases of the primordial
 binaries. The core radii at the halt can be explained by their energy
 budget.
\end{abstract}

\onecolumn

\section{Introduction}

Recently, observational informations concerning on the binary systems in
the globular cluster have been accumulated from photometric observations
of the eclipses (e.g. \cite{Mateo96}), spectroscopic observations
(e.g. \cite{Albrow+01}), low-mass X-ray binaries (see \cite{Liu+07}),
and also the color-magnitude diagrams
(e.g. \cite{Rubenstein+97}). \citet{Davis+08} also constrain the binary
fraction in NGC 6397 by the method of the color-magnitude diagrams, and
review the binary fractions in galactic globular clusters.

The binaries in the globular cluster play important roles for its
dynamical evolution, since the binaries work as energy source through
interactions with other stars or binaries (\cite{Heggie75}). Even if the
globular cluster has no binary at initial, the binaries are formed in due
course through three-body encounter, and energy generated from these
binaries halts core collapse of the globular cluster (\cite{Henon75}).

Furthermore, if the globular cluster contains a nonnegligible fraction
of primordial binaries, the dynamical evolution could be
affected. \citet{Goodman+89} (hereafter GH89) first showed theoretically
difference of evolutions with and without primordial binary. GH89
estimated that the core of the cluster with primordial binaries at the
halt of core contraction is larger than that of the cluster without
primordial binaries by order of magnitudes. GH89's estimate is based on
the model that the cluster core stop contracting when energy generated
by the primordial binaries is balanced with energy outflowing from the
inner region of the cluster to the outer region through two-body
relaxation. McMillan et al. (1990; 1991) first performed $N$-body
simulations of clusters with primordial binaries. They clearly showed
that the cluster cores stop contracting at larger cores than those
without primordial binaries. By means of Fokker-Planck model,
\citet{Gao+91} investigated post-collapse evolution of cluster with
the primordial binary. They showed that the cores of the clusters
continues to contract slowly after rapid core contraction and that the
gravothermal oscillations occur after the several ten half-mass
relaxation time of the slow core contraction.

In order to set a population of the primordial binaries in the cluster,
several parameters concerning the binary and its distributions, such as
mass fraction, and distribution of binding energies and eccentricities,
need to be specified. Although these parameters should be derived from
theories of star and cluster formations or be limited observationally,
sufficient informations have not yet been provided at present. Only a
few studies investigated the effect of these parameters on the
evolution. \citet{Heggie+92} performed $N$-body simulations of clusters
with two different mass fractions, $6$ and $12$ \%, of primordial
binaries, and \citet{Heggie+06} investigated the evolutions of clusters
with $0-100$ \% mass fraction of primordial binaries by means of
$N$-body simulations. \citet{Vesperini+94} (hereafter VC94) extended the
model of GH89, taking into account the core mass fraction of primordial
binaries and the distribution of the binding energies of the binaries.

In this paper, we focus on the dependence on binding energies of the
primordial binaries. By means of $N$-body simulations, we systematically
investigate the dynamical evolution of clusters with different initial
hardness of binding energies. We set the distribution of binding
energies of primordial binaries by delta function,
$\delta(x-E_{{\rm bin},0})$, where $E_{{\rm bin},0}$ is the initial
binding energies of the primordial binaries and
$E_{{\rm bin},0}=1,3,10,30,100,300kT_0$. Here, $1.5kT_0$ is the average
kinetic energy of stars in the cluster at the initial time. In previous
simulations, the distribution of the binding energy of the primordial
binaries are usually fixed at uniform distributions in
$\log E_{{\rm bin},0}$ (see \cite{McMillan+90}; \cite{Gao+91};
\cite{Heggie+92}; \cite{Heggie+06}).

We found that the evolutions of the cores are different according to the
distribution of the binding energies. If primordial binaries consist of
softer binaries, the binaries are disrupted before core collapse
through binary-single and binary-binary encounters, and do not affect
the core evolution. If primordial binaries consist of harder binaries,
the binaries escape from the cluster. This is because energy generated
by the harder binaries per one encounter is larger than the cluster
potential. They also do not affect the core evolution and the cluster
exhibits deep core collapse. In intermediate range, the binaries
efficiently heat the clusters, and the core collapses of the clusters
halt halfway. These behaviors are consistent with the theoretical
estimate by VC94.

These behaviors were not shown by previous works (\cite{McMillan+90};
\cite{Gao+91}; \cite{Heggie+92}; \cite{Heggie+06}; \cite{Fregeau+07};
\cite{Trenti+07}) who followed the dynamical evolution of clusters with
primordial binaries. This is mainly because the previous works fixed the
distributions of the binding energies of the primordial binaries at
uniform distributions in $\log E_{{\rm bin},0}$, and their fixed
distributions include binaries with hardness in soft, intermediate, and
hard ranges.

The structure of this paper is as follows. We describe simulation
methods in section \ref{sec:sim}. In section \ref{sec:results}, we
present results of simulations and investigate the effect of the
hardness of primordial binaries on the dynamical evolution of
clusters. In section \ref{sec:summary}, we summarize this paper.

\section{Simulation methods}
\label{sec:sim}

In this section, we describe simulation methods. In section
\ref{sec:IC}, we show initial conditions of cluster models. We perform
$N$-body simulations of the clusters by means of GORILLA, which is a
newly developed $N$-body simulation code for star clusters
(\cite{Tanikawa+09}, hereafter TF09). In section \ref{sec:gorilla}, we
outline GORILLA.

\subsection{Initial conditions}
\label{sec:IC}

By means of $N$-body simulations, we simulate the dynamical evolution of
$13$ clusters with point-mass particles and without an external tidal
field. The cluster models are shown in table
\ref{tab:initialmodel}. Each of $11$ clusters have equal-mass stars
and contains primordial binaries with equal binding energy. Among these
cluster models, binding energies of the primordial binaries, 
$E_{\rm bin,0}$, and mass fractions of the primordial binaries,
$f_{\rm b,0}$, are different, as shown in the second and third columns
of table \ref{tab:initialmodel}, respectively. The number of the
primordial binaries $N_{\rm b,0}$ is also shown in the fourth column of
table \ref{tab:initialmodel}. The remaining two clusters are reference
models as soft and hard limits; an $N=16384$ equal-mass cluster model
without primordial binaries, and a cluster model in which all binaries
are replaced by stars with double mass in the $f_{\rm b,0}=0.1$
models. The fifth and sixth columns in table \ref{tab:initialmodel} are
the mass fraction of the double mass stars, $f_{\rm d,0}$, and the
number of the double mass stars, $N_{\rm d,0}$, respectively.

In all models, we use Plummer's model to generate the initial
distribution of both single stars and center of mass of primordial
binaries in the clusters. The eccentricity distribution of the
primordial binaries is thermal distribution, $f(e)de=2ede$. The other
orbital elements of the primordial binaries, such as the inclination,
the longitude of the ascending node, and the argument of pericenter with
respect to the clusters, and the phase are distributed at random.

We adopt $N$-body standard units (\cite{Heggie+86}), such that
$G=M=-E=1$, where $G$ is the gravitational constant, $M$ is the total
mass of the cluster, $E$ is the total energy of the cluster not
including the internal binding energy of the primordial binaries. For
clusters with $16384$ stars, $1kT_0=1.0\times10^{-5}$.

\subsection{GORILLA: an $N$-body simulation code for globular clusters}
\label{sec:gorilla}

In order to perform $N$-body simulations of the above clusters, we use
an $N$-body simulation code for the star cluster, GORILLA (TF09). In
GORILLA, the orbits of cluster stars are integrated with a fourth-order
Hermite scheme with individual timestep (\cite{Makino+92}). The
timesteps of stars are quantized by power of $2$ (\cite{McMillan86})
in order to be adjusted to GRAPE, a special-purpose computer designed to
accelerate $N$-body simulations, where in our simulations we use
GRAPE-6/6A (\cite{Makino+03}; \cite{Fukushige+05}). Additionally, the
relative motions of two stars relatively isolated from other stars are
approximated as Kepler motions.

In order to determine timestep of each star in the fourth-order Hermite
scheme with individual timestep, we use the following criterion,
\begin{equation}
 \Delta t = \sqrt{\eta 
  \frac{\left|{\bf a}\right|\left|{\bf a}^{(2)}\right|+\left|{\bf
	 a}^{(1)}\right|^2}
  {\left|{\bf a}^{(1)}\right|\left|{\bf a}^{(3)}\right|+\left|{\bf
	 a}^{(2)}\right|^2}},
\end{equation}
where ${\bf a}$ is the acceleration of each star, ${\bf a}^{(n)}$ is
$n$-th order derivative of ${\bf a}$, and $\eta$ is an accuracy
parameter. We set the accuracy parameter $\eta=0.01$. We also use the
following timestep criterion for startup,
\begin{equation}
 \Delta t = \eta_{\rm s} \frac{\left|{\bf a}\right|}{\left|{\bf
				a}^{(1)}\right|},
\end{equation}
where $\eta_{\rm s}$ is a startup accuracy parameter. We set the startup
accuracy parameter $\eta_{\rm s}=0.0025$.

In GORILLA, the relative motions of binaries are approximated as Kepler
motion if the binary components (stars $k$ and $l$) satisfy either
of two conditions as follows.
\begin{description}
 \item [] Isolation conditions (A)
	     \begin{enumerate}
	      \item $|{\bf r}_{3}-{\bf r}_{{\rm
		    cm},kl}|>\alpha r_{{\rm apo},kl}$
	     \end{enumerate}
 \item [] Isolation conditions (B)
	     \begin{enumerate}
	      \item $e_{kl}>0.95$
	      \item $|{\bf r}_{3}-{\bf r}_{{\rm
		    cm},kl}|>\beta r_{{\rm rel},kl}$
	     \end{enumerate}
\end{description}
Here, ${\bf r}_3$ is the position of the nearest star originated
from the center of mass, ${\bf r}_{{\rm cm},kl}$, of stars $k$ and $l$,
$r_{{\rm apo}, kl}$ is the separation between
stars $k$ and $l$ at the apocenter, $r_{{\rm rel}, kl}$ is the
separation between stars $k$ and $l$ at a given time, and $e_{kl}$
is eccentricity.

\begin{figure}
 \begin{center}
  \FigureFile(140mm,90mm){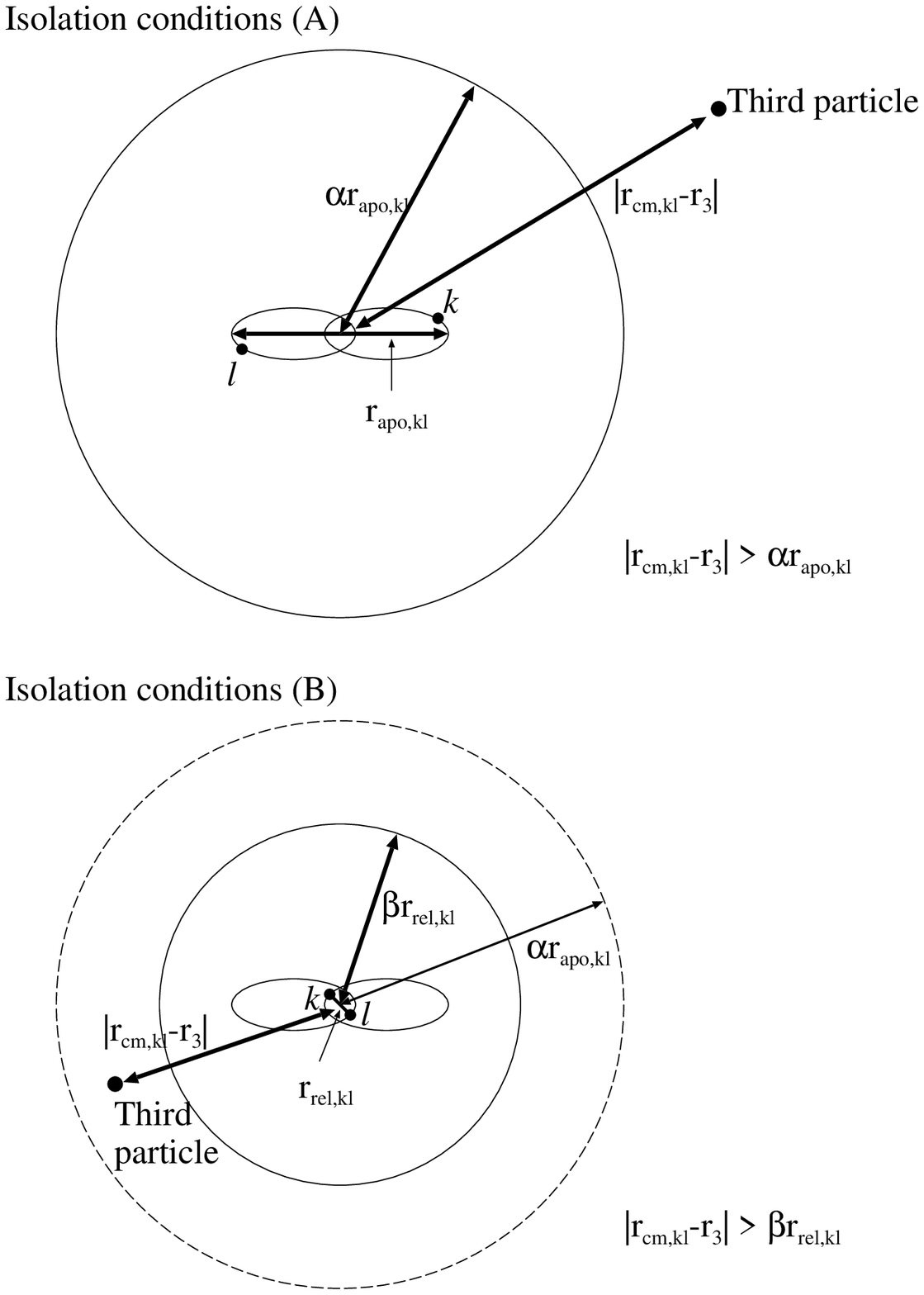}
  \caption{Illustration of binaries in isolation with conditions (A)
  (upper panel), and conditions (B) (lower panel).}
  \label{fig:alphabeta}
 \end{center}
\end{figure}

Condition (A) expresses whether the third star is separated enough
compared to apocentric distance of a binary. Condition (B) expresses
whether the third star is separated enough compared to the
instantaneous distance of the binary when the binary is highly
eccentric. The above conditions are illustrated in figure
\ref{fig:alphabeta}. Using dimensionless quantities $\alpha$ (called
apocentric parameter) and $\beta$ (called pericentric parameter), we
give criterions of the isolation. We show the apocentric and
pericentric parameters adopted in each cluster model in table
\ref{tab:parameters}, as well as the accuracy parameters, $\eta$ and
$\eta_{\rm s}$.

We set both the isolation conditions (A) and (B) in the following
reason. Consider a highly eccentric binary which does not satisfy
isolation conditions (A). Unless $r_{{\rm apo},kl}$ of the binary are
changed, the binary is not regarded as an isolated binary even if its
pericentric distance is very close. At the pericenter, the accuracy of
the orbital calculation is drastically decreased due to
$r_{{\rm rel},kl}$. Owing to isolation conditions (B), the binary is
regarded as an isolated binary around its pericenter. 

We apply the isolation not only to binaries, but also to unbound two
stars and hierarchical triple systems. Such unbound two stars are
sufficiently isolated from other stars and close to each other. If the
two stars satisfy the second condition of isolation conditions (B), they
are regarded as isolated stars. If both timesteps of the two stars are
less than $2^{-39}$ in standard $N$-body units, they are regarded as
close stars. In the hierarchical triple system, a binary and one star
orbit around each other. The binary in the hierarchical triple system is
called inner binary. If the inner binary is approximated as point mass,
we may regard the inner binary and the other star as a binary. Such
binary is called outer binary. If the hierarchical triple system is in
isolation, both the inner and outer binary satisfy isolation conditions
(A).

\section{Results}
\label{sec:results}

\subsection{Accuracy of $N$-body simulations}

\begin{figure}
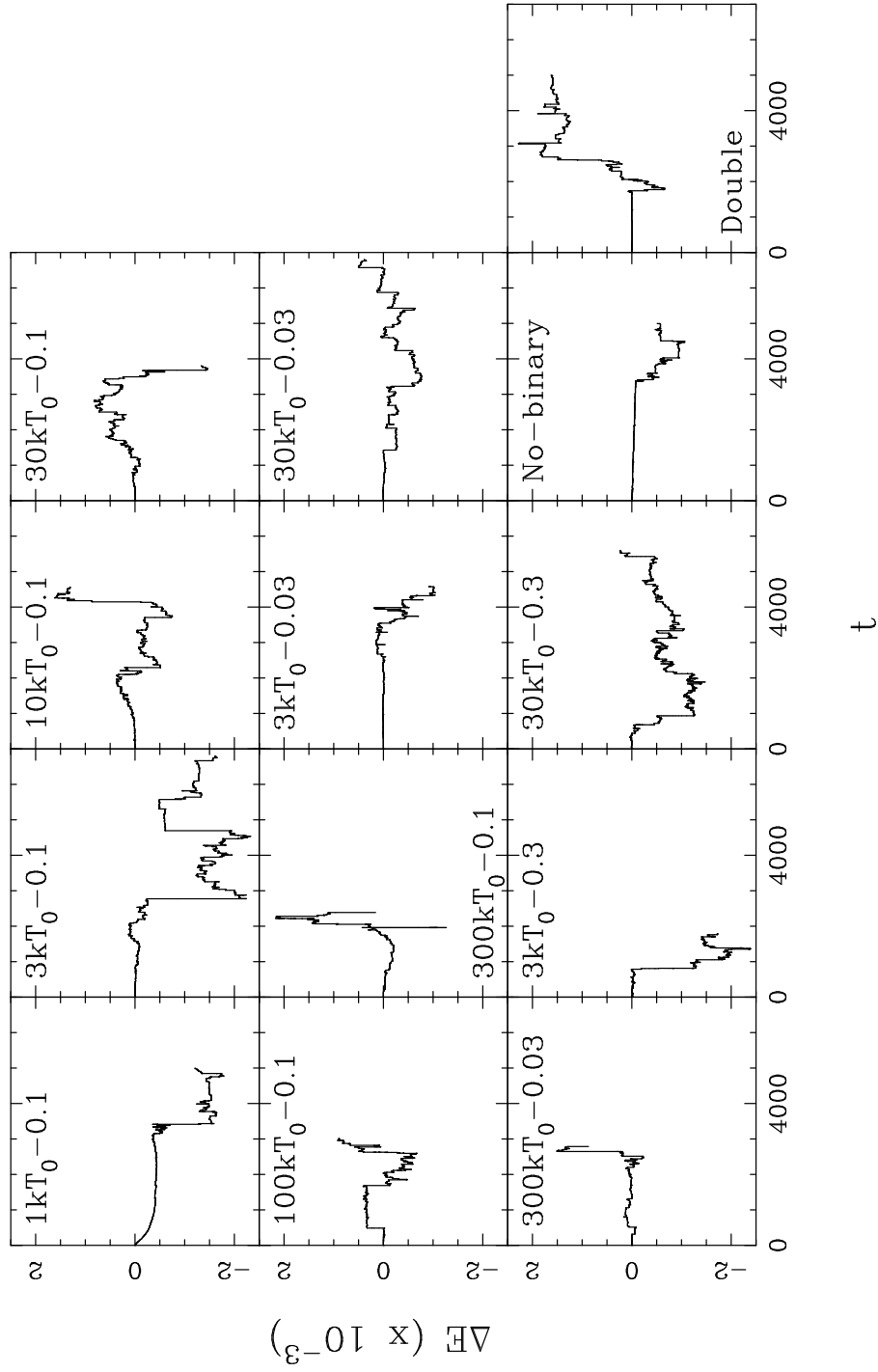

 \begin{center}
  \FigureFile(120mm,80mm){figure2.ps}
 \end{center}
 \caption{Energy errors as a function of simulation time.}
 \label{fig:err_all}
\end{figure}

Figure \ref{fig:err_all} shows energy errors as a function of simulation
time in all cluster models. They are all within $1$ \% ($\sim 0.0025$)
of the total energy not including the internal binding energy of the
primordial binaries at the initial time.

It seems that the energy errors $1$ \% are relatively large for studies
of the evolution of cluster cores whose energies are $1-10$ \% of the
total energies of the clusters, excluding the total binding energies of
the binaries. However, the energy errors do not much affect the core
evolutions. We found that large part of the energy errors attributes to the
binding energy of the binaries in the clusters when we use GORILLA
(TF09).

We also found that the energy errors do not affect properties of the
binaries, such as distribution of binding energies of the binaries. The
energy errors are at most $0.0025$ as seen in figure
\ref{fig:err_all}. On the other hand, the total binding energies of the
binaries increase by about $0.2$ for models No-binary and Double, and
all $f_{\rm b,0}$ models (see figure \ref{fig:energetics}). The energy
errors are only $\sim 1$ \% of the increased amount of the total binding
energy in each model.

\subsection{Core evolution and binary properties}
\label{sec:mainresults}

We first see the core radii, $r_{\rm c}$, and the half-mass
radii, $r_{\rm h}$, in the $f_{\rm b,0}=0.1$ models.
Figure \ref{fig:rc_all-t} shows the time evolution of the core radii
and half-mass radii of six $f_{\rm b,0}=0.1$ cluster models, and
models No-binary and Double. We calculate the core radii as in
\citet{Casertano+85} with the modifications described in
\citet{McMillan+90}. We calculate the core and half-mass radii at
each time unit, and average these radii over $10$ time units.

In models No-binary, $1kT_0-0.1$, and $3kT_0-0.1$, the clusters
experience deep core collapse, and gravothermal oscillations occur. The
core radii at the halts of the core collapse are $0.002 - 0.004$. In
models $10kT_0-0.1$, $30kT_0-0.1$, and $100kT_0-0.1$, the
core collapse stops halfway, and the cores contract
more slowly. The core radii at the halts of the core collapse are $0.05
- 0.1$. In models $300kT_0-0.1$ and Double, the clusters also experience
deep core collapse. In model Double, gravothermal oscillations
occur. The core radii at the halts of the core collapse are $0.005 -
0.02$. Among the clusters that experience deep core
collapse, the times when the core collapse stops and core bounce occurs
are different. In models No-binary and $1kT_0-0.1$, $t\sim3400$, in model
$3kT_0-0.1$, $t\sim4700$, and in models $300kT_0-0.1$ and Double,
$t\sim1700-2100$.

Next, we see the evolution of binary properties throughout the rest of
this subsection. Figure \ref{fig:energetics} shows in the thick curves
the increase of the total binding energy of the binaries, $\Delta E_{\rm
bin,tot}(t)$, in the $f_{\rm b,0}=0.1$ models and
models No-binary and Double. The arrows indicate the times when the core
collapse stops. The increase $\Delta E_{\rm bin,tot}(t)$ is given by
\begin{equation}
 \Delta E_{\rm bin,tot}(t) = \sum_{i}^{N_{\rm b}(t)} E_{{\rm bin},i} (t)
  - \sum_{i}^{N_{\rm b}(0)} E_{{\rm bin},i} (0),
\end{equation}
where $E_{{\rm bin},i}(t)$ is the binding energy of $i$-th binary at
time $t$, $N_{\rm b}(t)$ is the number of the binaries at time $t$
including binary escapers, and $\Delta E_{\rm bin,tot}(t)$ corresponds
to energy released by all the binaries. We can see that, in models
No-binary, $1kT_0-0.1$, and Double, the binaries do not release energy
until the core collapse stops. In model $3kT_0-0.1$, the binaries release
energy from $t \sim 1000$. In the other models, the binaries release
energy from $t\sim0$. 
After the core collapses in models No-binary, $1kT_0-0.1$, $3kT_0-0.1$,
and Double, three-body binaries release energy (discussed below).

Figure \ref{fig:energetics} shows in the thin curves the time evolution
of the total kinetic energy of escapers, $E_{\rm esc,tot}(t)$, in the
$f_{\rm b,0}=0.1$ models and models No-binary and Double. The escapers
are defined as stars, regardless of single stars, binaries, or
hierarchical triple systems, satisfying both conditions as follows.
\begin{description}
 \item [(a)] The sum of the kinetic and potential energy of the single
	     star (or the center of mass of the binary or hierarchical
	     triple system) is positive.
 \item [(b)] The distance between the star and the center of the cluster
	     is more than $40$ length units.
\end{description}

We can see that, in models No-binary, $1kT_0-0.1$, and Double, the total
energy of the escapers is small just before the halts of the core
collapse, such that $E_{\rm esc,tot}(t) \sim 1 \times 10^{-3}$. In model
$300kT_0-0.1$, the thin curve is almost overlapped with the thick curve,
which indicates that the escapers carry away almost all energy released
by the binaries. In the other models, $E_{\rm esc,tot}(t)$ at the halts
of core collapse is larger than those of models No-binary, $1kT_0-0.1$,
and Double by an order of magnitude.

Figure \ref{fig:binnum} shows the time evolution of the number of
binaries, $N_{\rm b}$, in the $f_{\rm b,0}=0.1$ models and models
No-binary and Double. For model Double, the number of the double mass
stars is also plotted. In each panel, the thick curve shows the number
of binaries (or double mass stars) within the cluster, and the thin
curve shows the total number of binaries (or double mass stars)
including escapers. The arrows indicate the times when the core collapse
stops.

We can see that, in model No-binary, the binaries increase after deep
core collapse. These binaries are the three-body binaries. In models
$1kT_0-0.1$ and $3kT_0-0.1$, the numbers of the binaries rapidly decrease
before deep core collapse. After deep core collapse, the total numbers
of the binaries including binary escapers increase. In these models, the
three-body binaries are also formed. In models $10kT_0-0.1$, $30kT_0-0.1$,
$100kT_0-0.1$, and $300kT_0-0.1$, the numbers of the binaries
monotonically decrease. In models No-binary, $1kT_0-0.1$, and
$3kT_0-0.1$, the three-body binaries are formed after deep core
collapse. When the simulations are finished, the numbers of the
three-body binaries are $29$, $23$, and $15$, and the numbers of the
escapers of the three-body binaries are $22$, $19$, and $15$ in models
No-binary, $1kT_0-0.1$, and $3kT_0-0.1$, respectively. In model Double,
the three-body binaries are also formed after deep core collapse. The
total number of these binaries including the escapers is $27$ at
$t=5000$. The number of the binaries composed of the two double mass
stars is $25$. Two binaries are composed of one double and one single
mass stars. Nearly all binaries are composed of the double mass
stars. The number of the binaries within the cluster is $2$, both of
which are composed of the two double mass stars. In this model, the
number of the double mass stars monotonically decreases.

\begin{figure}
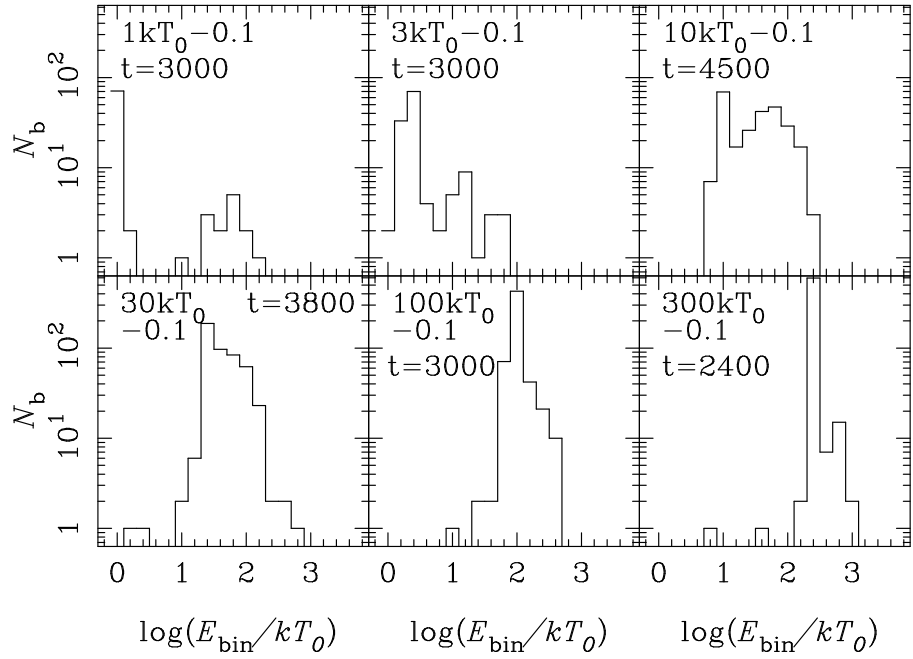

 \begin{center}
  \FigureFile(120mm,100mm){figure6.ps}
 \end{center}
 \caption{The number of binaries, $N_{\rm b}$, in each logarithmic bin
 of the binding energy, $E_{\rm bin}$.}
\label{fig:edis}
\end{figure}

Figure \ref{fig:edis} shows the number of binaries ($N_{\rm b}$) in each
logarithmic bin of binding energies of the binaries ($E_{\rm bin}$) for
the $f_{\rm b,0}=0.1$ models at the time indicated in each panel. All
the binaries in the clusters at those time are counted, and are
primordial binaries. There are no three-body binaries. In all models,
the peaks around the initial binding energies, $E_{\rm bin}$, can be
seen. The distributions of the binding energy spread towards the larger
sides.

\begin{figure}
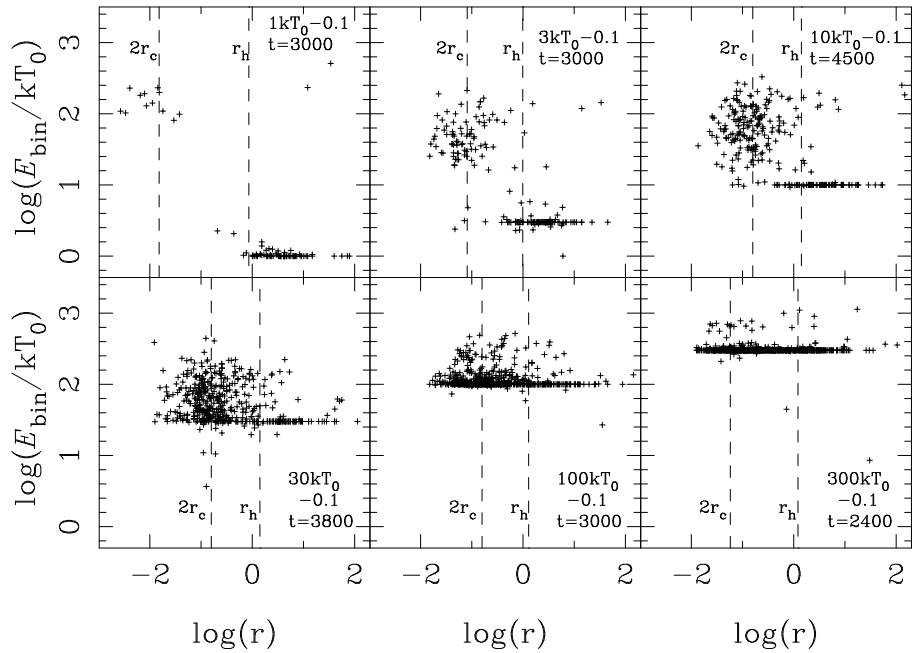

 \begin{center}
  \FigureFile(120mm,100mm){figure7.ps}
 \end{center}
 \caption{Binding energies as a distance from cluster center of each
 binary  at the time indicated in the panels. The dashed lines show the
 half-mass radii and twice the core radii at the time.}
\label{fig:x-ebin}
\end{figure}

In figure \ref{fig:x-ebin}, the binding energies are shown as a distance
from of the cluster center for all the binaries in models $1kT_0-0.1$,
$3kT_0-0.1$, $10kT_0-0.1$, $30kT_0-0.1$, $100kT_0-0.1$, and
$300kT_0-0.1$ at the time indicated in the panels. All the binaries in
the clusters at those time are counted, and are primordial binaries. The
dashed lines show the half-mass radii and twice the core radii at the
time. Within the half-mass radii, the distributions of the binding
energies are greatly changed from those at the initial time in all the
models except model $300kT_0-0.1$. The distributions of the binding
energies in these models are similar. The distributions center on about
$100kT_0$, and range from more than $10kT_0$ to $300kT_0$. On the other
hand, outside the half-mass radii of these models, the distributions of
the binding energies are little changed from the initial time. In model
$300kT_0-0.1$, even inside the half-mass radii, the distribution of the
binding energies is not changed so much.

In model $1kT_0-0.1$, there are little binaries between $2r_{\rm c}$ and
$r_{\rm h}$. The binaries do not increase between $2r_{\rm c}$ and
$r_{\rm h}$ despite of mass segregation, unlike the core. Furthermore,
the binaries are disrupted through binary-single encounters even between
$2r_{\rm c}$ and $r_{\rm h}$. The large semi-major axes of the $1kT_0$
binaries result in frequent binary-single encounters despite of stellar
density in this region smaller than that in the core, and the softness
of the binaries results in the destruction of the binaries through the
encounters despite of the average stellar kinetic energy in this region
smaller than that in the core.

Figure \ref{fig:binfrac} shows the time evolution of the mass fraction,
$f_{\rm b}$, of the binaries inside the core radii (upper curves) and
half-mass radii (lower curves) in the $f_{\rm b,0}=0.1$ models and model
No-binary. For model Double, mass fraction, $f_{\rm d}$, of the double
mass stars are shown. In model $1kT_0-0.1$, both the fractions inside
the core and half-mass radii decrease from $0.1$ to $0.02$ until the
deep core collapse occurs. In model $3kT_0-0.1$, the fraction inside the
core radius increases up to $0.3$ at $t\sim1000$, and decreases down to
$\sim 0.04$ at the time when the deep core collapse occurs. In models
$10kT_0-0.1$, $30kT_0-0.1$, $100kT_0-0.1$, $300kT_0-0.1$, and Double,
the fraction inside the core radii increases until the core collapse
stops. After that time, the fractions stop increasing.

\begin{figure}
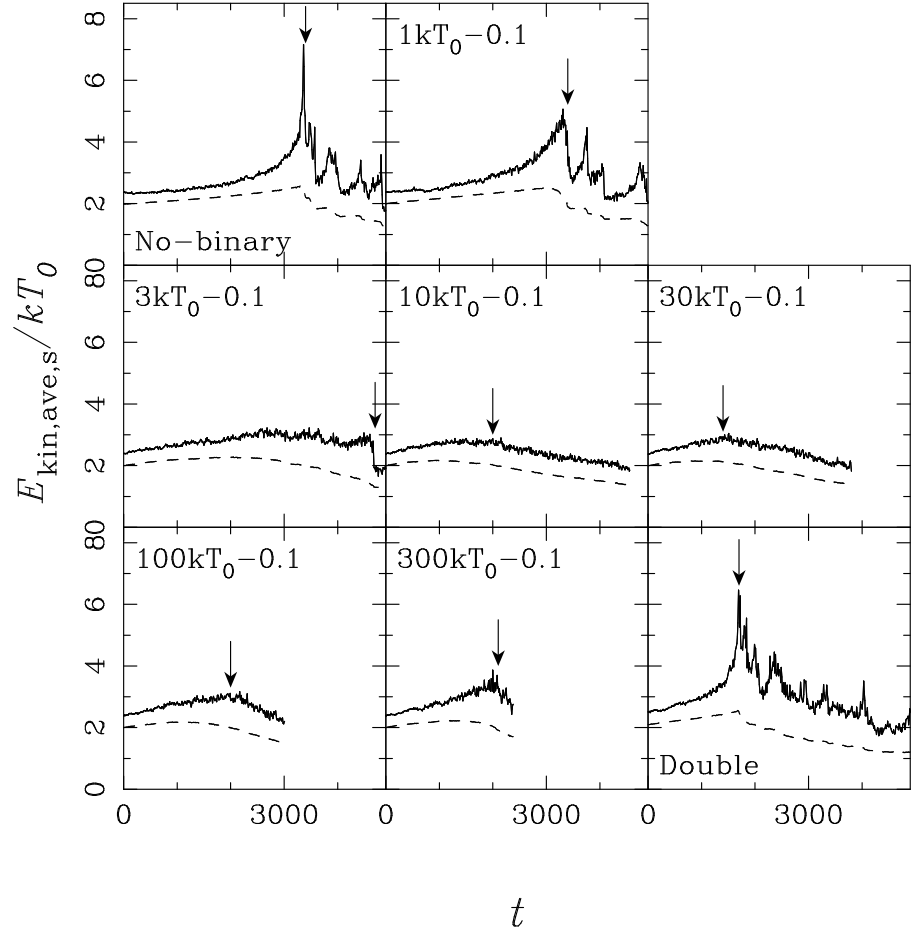

\begin{center}
\FigureFile(120mm,100mm){figure9.ps}
\end{center}
\caption{Time evolution of the mean kinetic energy of the single
 stars inside the core radii (solid curves) and the half-mass radii
 (dashed curves) in the $f_{\rm b,0}=0.1$ models and models No-binary
 and Double. The arrows show the time when the core collapse stops.}
\label{fig:ekinsin}
\end{figure}

Figure \ref{fig:ekinsin} shows the time evolution of the mean kinetic
energies of the single stars, $E_{\rm kin,ave,s}$, inside the core radii
(solid curves) and the half-mass radii (dashed curves) in the
$f_{\rm b,0}=0.1$ models and models No-binary and Double. In models
No-binary, $1kT_0-0.1$, $300kT_0-0.1$, and Double, at the moment of core
collapse, the mean kinetic energies of  the single stars in the cores
increase several times more than those at the initial time. On the
other hand, they nearly keep constant during the evolution in models
$3kT_0-0.1$, $10kT_0-0.1$, $30kT_0-0.1$, and $100kT_0-0.1$.

Figure \ref{fig:ekinbin} shows the time evolution of the mean kinetic
energy of the binaries, $E_{\rm kin,ave,b}$, inside the core radii in
the $f_{\rm b,0}=0.1$ models. We do not
show those inside the half-mass radii, since they are nearly the
same as inside the core radii. The fluctuation is large in model
$1kT_0-0.1$, since the number of the binaries is small. Except model
$1kT_0-0.1$,
the mean kinetic energy of the binaries in the core keep nearly constant
during the evolution. In model $1kT_0-0.1$, the mean kinetic energy
largely increases at the deep core collapse.

\subsection{Interpretation}
\label{sec:interpretation}

On both softer (models No-binary, $1kT_0-0.1$, and $3kT_0-0.1$) and harder
(models $300kT_0-0.1$ and Double) hardness, the clusters undergo deep
core collapse. On the other
hand, in the intermediate hardness (models $10kT_0-0.1$, $30kT_0-0.1$,
and $100kT_0-0.1$), the clusters exhibit shallower core collapse. The
depth of the core collapse depends on the amount of energy heating core
generated by the primordial binaries. The larger the amount of energy
is, the shallower core collapse becomes, and vice versa.

The amount of the energy heating the core depends on whether the
primordial binaries become harder or not through binary-single and
binary-binary encounters, and whether the single stars and binaries
heated by such encounters are ejected or not from the clusters. Whether
the binaries become harder or not depends on whether the binding energy
of the binaries is larger or not than a critical energy $E_{\rm
crit,H}$, which, in this case, corresponds to the average kinetic energy
of the surrounding single stars, $E_{\rm kin,ave,s}$ (Heggie's law:
\cite{Heggie75}). If
$E_{\rm bin}\lesssim E_{\rm crit,H}\sim E_{\rm kin,ave,s}$, the binaries
are on average destroyed through binary-single encounters. Therefore,
the binaries cannot heat the core. If $E_{\rm bin}\gtrsim E_{\rm
crit,H}$, the binaries become harder and harder through series of
binary-single encounters, and then heat the core.
Since most binary-single encounters occur in the core, we should adopt
the average kinetic energy of the single stars in the core.
In Plummer's model, which is the initial condition of our models, the
average kinetic energy of the single stars
in the core
is about $2.5kT_0$, which can be
seen in figure \ref{fig:ekinsin}. Therefore,
\begin{equation}
E_{\rm crit,H} \sim E_{\rm kin,ave,s} \sim 2.5kT,
\end{equation}
where $1.5kT$ is the average stellar kinetic energy in the whole cluster
at a given time.

Whether the single stars and binaries heated through encounters are
ejected or not depends on whether the kinetic energies transformed from
the binding energies of binaries are larger or not than the potential
depth of the whole cluster. The increase of the binding energy at single
binary-single encounter, $\Delta E_{\rm bin}$, is on average  $\Delta
E_{\rm bin} \simeq 0.4 E_{\rm bin}$ (\cite{Heggie75}). According to
conservation of momentum, two thirds of the energy released by the
binaries go to the single star, and the rest goes to the binary, on
average. The condition to eject both the single star and binary is
$\Delta E_{\rm bin}/3>2m|\Phi_{\rm c}|$, where $m$ is the mass of the
single mass stars, and $\Phi_{\rm c}$ is the potential of the core. If
we define a critical energy for ejection as $E_{\rm crit,E}$,
$0.4E_{\rm crit,E}/3>2m|\Phi_{\rm c}|$. In Plummer's model,
$m|\Phi_{\rm c}| \simeq 10kT$. Therefore, the ejection occurs when
\begin{equation}
 E_{\rm bin} > E_{\rm crit,E} \sim \frac{6}{0.4} m|\Phi_{\rm c}| =
  150kT.
  \label{eq:ejection}
\end{equation}

\begin{figure}
\begin{center}
\FigureFile(120mm,100mm){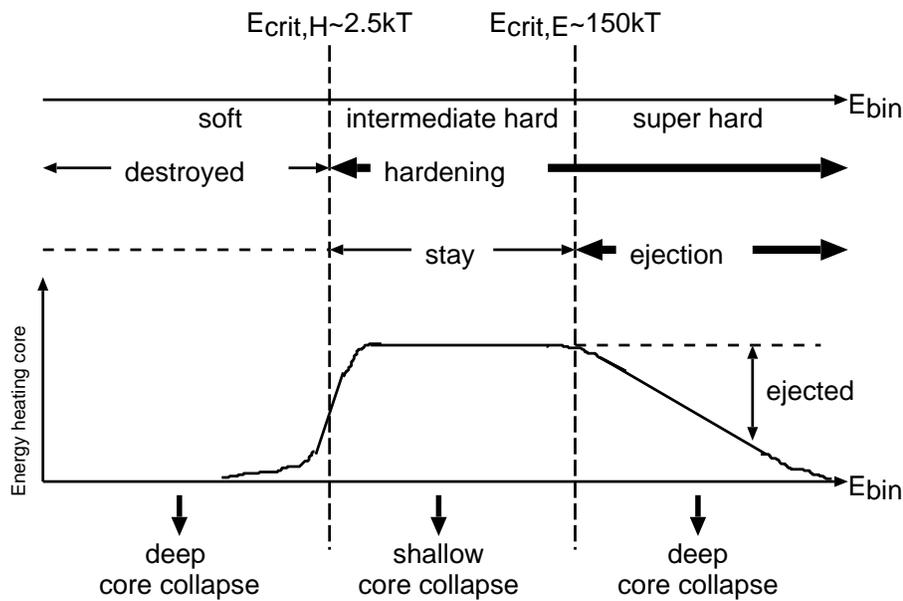}
\end{center}
\caption{Interpretation of core evolution.}
\label{fig:mechanism}
\end{figure}

In summary, whether primordial binaries can heat the core or not are
different among three ranges of hardness divided by two critical
hardness $E_{\rm crit,H}$ and $E_{\rm crit,E}$; (a) $E_{\rm bin}\lesssim
E_{\rm crit,H}$, (b) $E_{\rm crit,H} \lesssim E_{\rm bin} \lesssim
E_{\rm crit,E}$, and (c) $E_{\rm bin} \gtrsim E_{\rm crit,E}$, which are
illustrated in figure \ref{fig:mechanism}. Here and hereafter, binaries
with smaller binding energy than $E_{\rm crit,H}$ are called soft, and
those with larger binding energy than $E_{\rm crit,H}$ are called
hard. If $E_{\rm bin} \lesssim E_{\rm crit,H}$ (soft range, (a)), the
primordial binaries are destroyed through encounters, and can not heat
the core. If
$E_{\rm crit,H}\lesssim E_{\rm bin} \lesssim E_{\rm crit,E}$
(intermediate hard range, (b)), the primordial binaries become harder
and harder and continually heat the core. If
$E_{\rm bin}\gtrsim E_{\rm crit, E}$ (super hard range, (c)), the
primordial binaries release their binding energies to the surrounding
single stars and binaries, but the single stars and binaries are ejected
from the cluster, and then the primordial binaries can not so much heat
the core. In the following sub-subsections, we see the simulation
results for each range of hardness.

\subsubsection{Soft range}

\begin{figure}
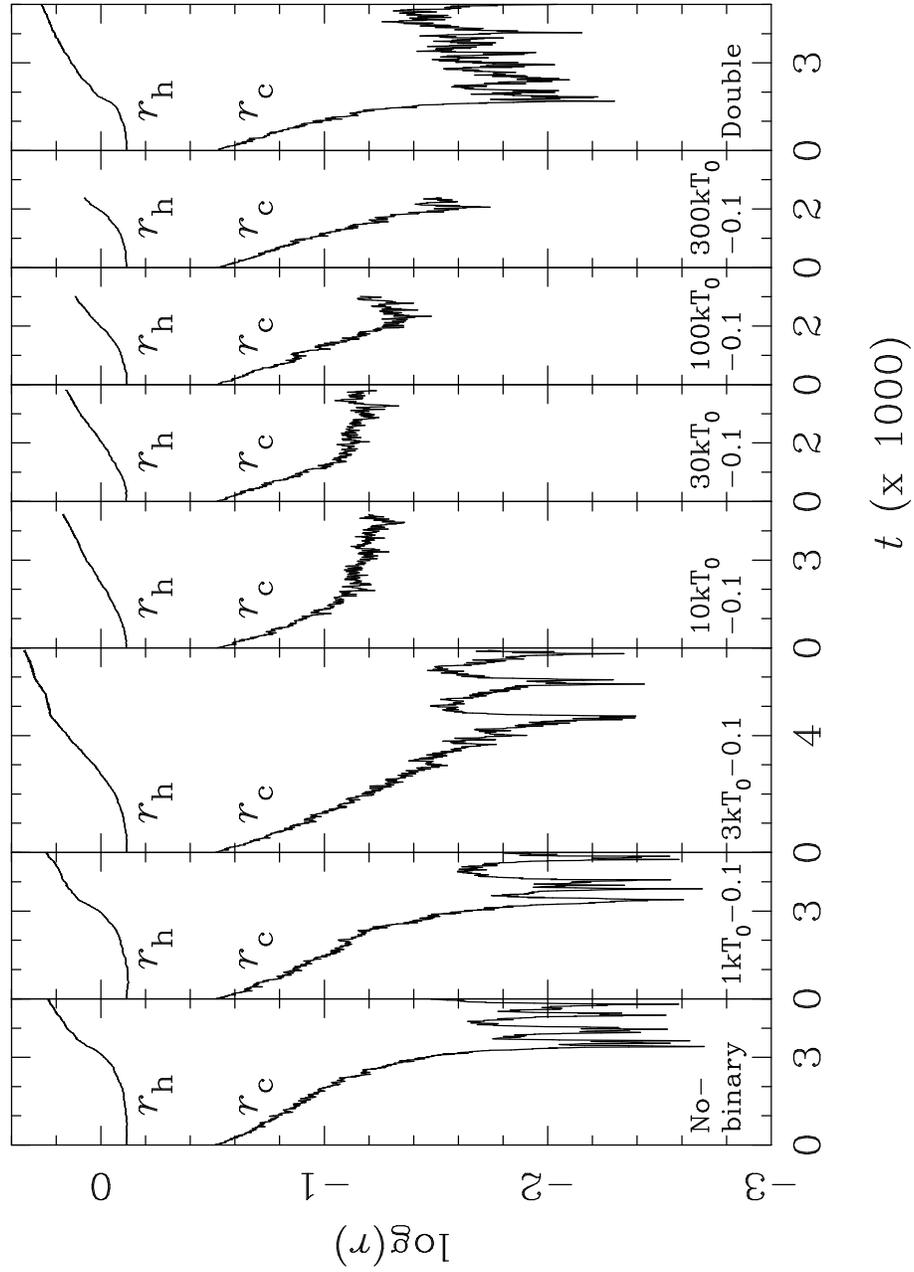

\begin{center}
\FigureFile(120mm,80mm){figure3.ps}
\end{center}
\caption{Time evolution of the core radii, $r_{\rm c}$, and half-mass
 radii, $r_{\rm h}$, of $f_{\rm b,0}=0.1$ cluster models, and models
 No-binary and Double.}
\label{fig:rc_all-t}
\end{figure}

\begin{figure}
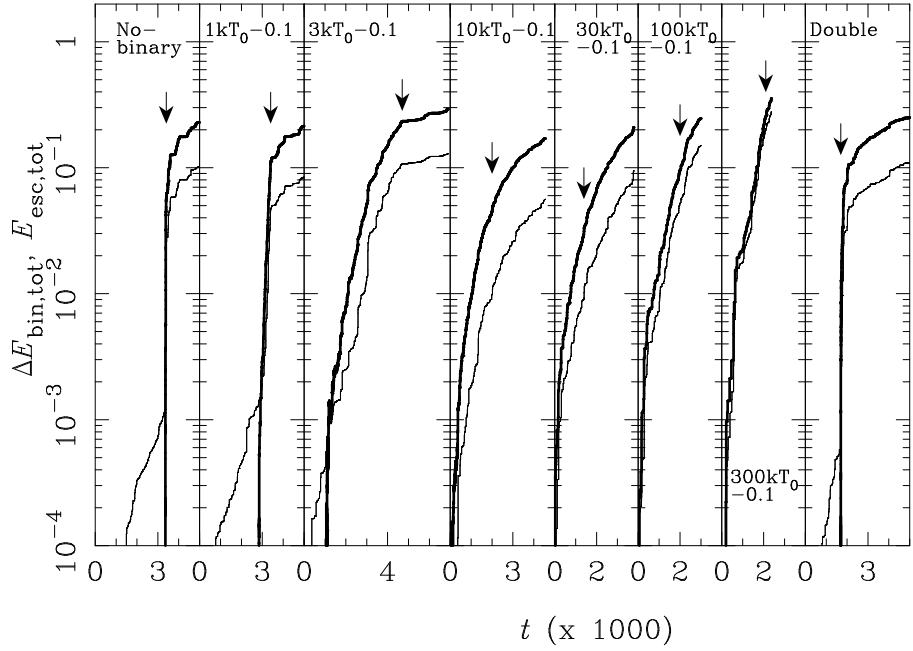

\begin{center}
\FigureFile(120mm,100mm){figure4.ps}
\end{center}
 \caption{The increase of the total binding energy of the binaries,
 $\Delta E_{\rm bin,tot}(t)$ (thick curves), which corresponds to
 energy generated by the binaries, and the total kinetic energy of
 escapers, $E_{\rm esc,tot}(t)$, from the clusters (thin curves). The
 arrows indicate the times when the core collapse stops.}
 \label{fig:energetics}
\end{figure}

\begin{figure}
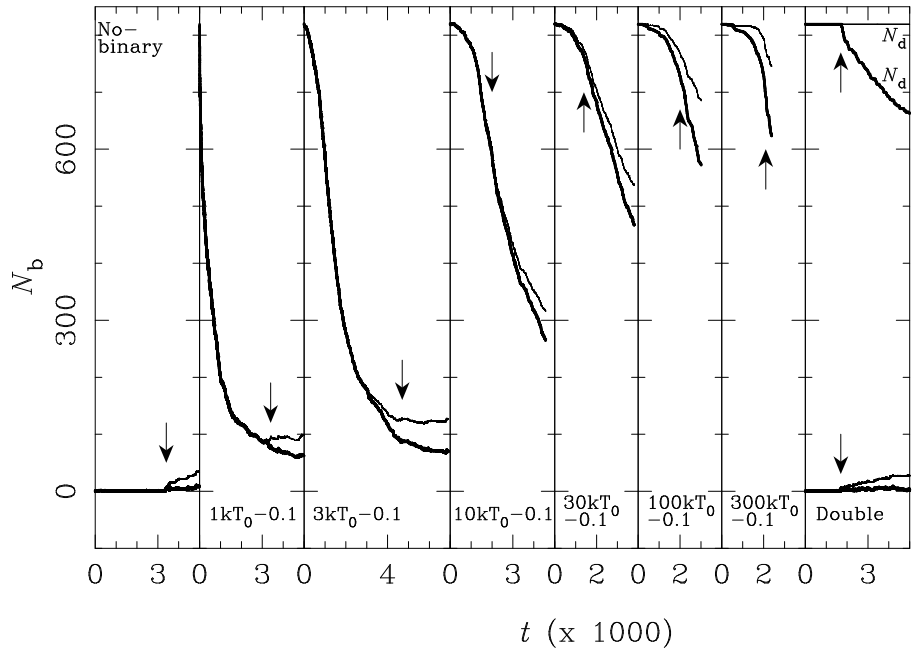

 \begin{center}
  \FigureFile(120mm,100mm){figure5.ps}
 \end{center}
 \caption{Time evolution of the number of binaries, $N_{\rm b}$, in
 the $f_{\rm b,0}=0.1$ models and models No-binary and Double. For
 model Double, the
 number of the double mass stars is also plotted. The thick curves
 indicate the numbers of binaries (or double mass stars) within the
 clusters, and the thin curves indicate the total numbers of binaries,
 (or double mass stars) including escapers. The arrows indicate the
 times when the core collapse stops.}
\label{fig:binnum}
\end{figure}

In the soft range, $E_{\rm bin}\lesssim E_{\rm crit,H}(\sim 2.5kT)$, the
primordial binaries are destroyed through encounters, and cannot heat
the core. We can see, in figure \ref{fig:binnum} (the second left
panel), the number of binaries for model $1kT_0-0.1$ rapidly decreases,
and, in figure \ref{fig:energetics} (the second left panel), primordial
binaries do not release energy from the beginning. After then, their
evolutions are almost identical to the case without primordial binaries,
and lead to deep core collapse. Figure \ref{fig:rc_all-t} and
\ref{fig:energetics} (the first and second left panels) show that the
core evolution and energy generations are very similar between models
No-binary and $1kT_0-0.1$.

Model $3kT_0-0.1$ exhibits a mixed behavior between soft and
intermediate hard ranges. While the number of the binaries rapidly
decreases, shown in figure \ref{fig:binnum} (the third left panel), the
binaries continually release energy, shown in figure
\ref{fig:energetics} (the third left panel). A population whose binding
energy is harder than initial value can be seen in figure \ref{fig:edis}
(the second left upper panel). Although that heating is not so large to
stop core collapse, it makes the time to the core collapse longer,
compared to the case without primordial binaries, shown in figure
\ref{fig:rc_all-t} (the first, second, and third left panels).
Note that at the very beginning the mass fraction of binary in the core
increase, shown in figure \ref{fig:binfrac} (the third left panel),
due to mass segregation of single star and binary, and sinking of binary
to the core. After $t=1000$, the mass fraction of binary in the core
turns to decreasing, since the binaries are destroyed in the core.

\subsubsection{Intermediate hard range} 

In the intermediate range, $E_{\rm crit,H} (\sim 2.5kT) \lesssim E_{\rm
bin} \lesssim E_{\rm crit,E} (\sim 150kT)$, the primordial binaries
continually heat the core, and becomes harder and harder. We can see in
figure \ref{fig:energetics} (the fourth, fifth, and sixth left panels)
the
primordial binaries release the binding energy, and in figure
\ref{fig:edis} (panels $10kT_0-0.1$, $30kT_0-0.1$, and $100kT_0-0.1$),
populations whose binding energy become harder from initial values can
be seen. The continually released heat halts the core collapse halfway.

\subsubsection{Super hard range}

In the super hard range, $E_{\rm bin} \gtrsim E_{\rm crit,E} (\sim
150kT)$, the kinetic energy transformed from binding energy of
primordial binaries through encounters is so large to be ejected from
whole cluster immediately, and the primordial binaries cannot heat the
core. Figure \ref{fig:energetics} (the second right panels)
shows the released binding energy in total (in the thick curves), and
those of escapers (in the thin curve) are very close in model
$300kT_0-0.1$, which means almost all released energy from the
primordial binaries is brought away from the cluster by the
escapers. Therefore, the primordial binaries can neither heat the core,
nor stop core collapse. Their evolutions become similar to the case in
which binaries are replaced by the double mass stars, which is shown in
figure \ref{fig:rc_all-t} (first and second right panels).

\subsubsection{Theoretical estimate}
\label{sec:formula}

We theoretically estimate core size at the halt of core contraction,
assuming that the energy outflowing from the inner region of the cluster
to the outer region at each unit time, $dE_{\rm h}/dt$, is balanced with
the energy provided for the core through binary interactions at each
unit time, $dE_{\rm c}/dt$, such as
\begin{equation}
 \frac{dE_{\rm h}}{dt} = \frac{dE_{\rm c}}{dt}.
  \label{eq:balance}
\end{equation}
The argument here and hereafter is based on VC94.

The energy outflowing from the inner region of the cluster to the outer
region at each unit time, $dE_{\rm h}/dt$, is given by
\begin{equation}
 \frac{dE_{\rm h}}{dt} = \frac{|E|}{\gamma t_{\rm rh}} \approx
  \frac{0.2}{\gamma} \frac{GM^2}{t_{\rm rh}r_{\rm h}},
  \label{eq:eh}
\end{equation}
where $G$ is the gravitational constant, $M$ is the total mass of the
cluster, and $\gamma$ is a numerical coefficient relating the energy
outflow rate to the half-mass relaxation time. The half-mass relaxation
time $t_{\rm rh}$ is expressed as
\begin{equation} 
 t_{\rm rh} = \frac{0.138M^{1/2}{r_{\rm
  h}}^{3/2}}{G^{1/2}\bar{m}\log(0.4N)},
  \label{eq:trh}
\end{equation}
where and $\bar{m}$ is the average mass of its stars
(\cite{Spitzer87}). We use $|E| \sim 0.2GM^2/r_{\rm h}$, which comes
from virial theorem, $|E| \sim GM^2/4r_{\rm v}$, where $r_{\rm v}$ is
the virial radius, and $r_{\rm h} \sim 0.8r_{\rm v}$ in Plummer's
model.

The energy provided for the core by binary interactions at each unit
time, $dE_{\rm c}/dt$, is expressed as
\begin{equation}
 \frac{dE_{\rm c}}{dt} = \lambda
  V_{\rm c} \frac{G^2m^3}{v_{\rm s,c}} {n_{\rm c}}^2 
  \left(A_{\rm bs} + A_{\rm bb} \right)
 \label{eq:ec}
\end{equation}
where $V_{\rm c}$ is the core volume, $v_{\rm s,c}$ is one dimensional
velocity dispersion of single stars in the core, $n_{\rm c}$ is the
total number density of the single stars and binaries in the core, and
$A_{\rm bs}$ and $A_{\rm bb}$ are, respectively, dimensionless
efficiency factors for energy provided for a cluster through
binary-single and binary-binary encounters. The core volume,
$V_{\rm c}$, is given by $V_{\rm c} = (4\pi/3) r_{\rm c}^3$. We
introduce a free parameter $\lambda$, which is nearly equal to unity,
and actually we adopt $\lambda=0.45$ as discussed later. The free
parameter $\lambda$ is the fraction of the core volume where most
energy is released through binary-single and binary-binary encounters
in the unit of the core volume.

Substituting equations (\ref{eq:eh}) and (\ref{eq:ec}) into equation
(\ref{eq:balance}), we express the ratio of the core to half-mass radii
at the halt of core collapse as
\begin{equation}
 \frac{r_{\rm c}}{r_{\rm h}} = \frac{0.0196\lambda}{\log_{10}(0.4N)}
  \left( \frac{v_{\rm s,c}}{v_{\rm h}}\right)^3
  \left( \frac{\gamma}{10} \right)
  (2-f_{\rm b,c})^4 \left(A_{\rm bs} + A_{\rm bb}\right)
  \label{eq:theory}
\end{equation}
where $f_{\rm b,c}$ is mass fraction of binaries in the core,
$v_{\rm h}$ is the one dimensional half-mass velocity dispersion,
and $|E| \sim 3M{v_{\rm h}}^2/2$ {\it i.e.}
$3{v_{\rm h}}^2/2 \sim GM/5r_{\rm h}$. Here, we define the core radius
as
\begin{equation}
 r_{\rm c} = \sqrt{\frac{9{v_{\rm c}}^2}{4\pi G \rho_{\rm c}}} =
  \sqrt{\frac{9{v_{\rm s,c}}^2(2-f_{\rm b,c})^2}{16\pi G mn_{\rm c}}},
  \label{eq:rc}
\end{equation}
where ${v_{\rm c}}^2={v_{\rm s,c}}^2(2-f_{\rm b,c})/2$ and $\rho_{\rm
c}=2mn_{\rm c}/(2-f_{\rm b,c})$ are, respectively, the average velocity
dispersion and mass density in the core.

The dimensionless efficiency factors of binary-single and binary-binary
encounters are indicated by $A_{\rm bs}$ and $A_{\rm bb}$. We derive
these factors below, although the derivations are nearly the same way as
VC94. This is because two points are different from theirs. One is the
unit of energy. In this paper, $kT$ indicates the average stellar
kinetic energy in the whole cluster. In VC94, however, $kT$ indicates
the average stellar kinetic energy in the cluster core. The other is
the numerical factor of the cross section of binary-binary encounters,
$S$, which is first seen in equation (\ref{eq:sigmabb}).

The dimensionless efficiency factors are, respectively, expressed as
\begin{equation}
 A_{\rm bs} = 
  \left[\frac{2(1-f_{\rm b,c})}{2-f_{\rm b,c}}\right]\left(\frac{f_{\rm
   b,c}}{2-f_{\rm b,c}}\right) \int f(x) \left[g(x) h(x) + g'(x) \right]
   dx,
   \label{eq:Abs}
\end{equation}
and
\begin{equation}
 A_{\rm bb} = \frac{1}{2}\left(\frac{f_{\rm b,c}}{2-f_{\rm
			  b,c}}\right)^2 \int
 f(x_1)f(x_2)G(x_1,x_2)H(x_1,x_2) dx_1 dx_2,
   \label{eq:Abb}
\end{equation}
where $x$, $x_1$, and $x_2$ are the binding energy of the binaries in
the unit of $kT$. The function $f(x)$ is the distribution function of
the binding energies of the binaries in the core. The functions $g(x)$,
$g'(x)$ and $G(x_1,x_2)$ are, respectively, the dimensionless hardening
rates of the binary with the binding energy $x$ which is not destroyed
in a sea of single stars, the binary with the binding energy $x$ which
is destroyed in a sea of single stars, and the binary with the binding
energy $x_1$ in a sea of binaries with the binding energy $x_2$. The
functions $h(x)$ and $H(x_1,x_2)$ are, respectively, the efficiency
ratios of heating of the core to hardening of the binaries at each
interaction between a single star and a binary with the binding energy
$x$, and that between binaries with the binding energies $x_1$, and
$x_2$. The $h(x)$ and $H(x_1,x_2)$ become less than unity when single
stars and binaries involved with encounters are ejected from the cluster
immediately after the encounters. When the binaries are destroyed
through binary-single encounters, $h(x)=1$, since they are hardly
ejected.

The dimensionless hardening rate $g(x)$ is expressed as
\begin{equation}
 g(x) = 1.66 \left(\frac{x}{C}\right)^{-1} \int R_{\rm bs} (x/C,\Delta)
  \Delta d\Delta,
\end{equation}
where $R_{\rm bs}$ is the dimensionless rate of the interactions that
the binary with the binding energy $x$ hardens to the binding energy
$(1+\Delta)x$ in a sea of single stars, and $C(=1.7)$ is a correction
factor in order to set the unit of the binding energy to be
$kT_{\rm c}(=1.7kT)$, which is one dimensional kinetic energy of single
stars in the core. The integral of the dimensionless rate $R_{\rm bs}$
over $\Delta$ is described in equation (49) of \citet{Heggie+93},
and obtained as a function of the binding energy of the binary in the
unit of $kT_{\rm c}$. 

The dimensionless hardening rate $g'(x)$ is expressed as
\begin{equation}
 g'(x) = - 1.66 \left(\frac{x}{C}\right)^{-1} R'_{\rm bs} (x/C),
\end{equation}
where $R'_{\rm bs}$ is the dimensionless rate of the interactions that
the binary with the binding energy $x$ is destroyed in a sea of single
stars. The dimensionless rate $R'_{\rm bs}$ is described in equation
(5.12) of \citet{Hut+83}, and also obtained as a function of the
binding energy of the binary in the unit of $kT_{\rm c}$.

The dimensionless hardening rate $G(x_1,x_2)$ is expressed as
\begin{equation}
 G(x_1,x_2) = 1.66 \left(\frac{x_1+x_2}{C}\right)^{-1} R_{\rm bb}
  (x_1,x_2) \Delta,
\end{equation}
where $R_{\rm bb}$ is the dimensionless rate of the interaction that the
binary with the binding energies $x_1$ hardens to the binding energy
$(1+\Delta)(x_1+x_2)$ in a sea of binaries with the binding energy $x_2$
when $x_1>x_2$, and the binaries with the binding energy $x_2$ are
destroyed. We consider only binary-binary interaction which results in
the destruction of the softer binary, since Mikkola (1983a; 1983b;
1984a; 1984b) showed that the binary-binary interaction not involving
the destruction of one binary has small contribution to the heating of
the cluster. The dimensionless rate is averaged over $\Delta$, since the
number of binary-binary scattering experiments is much smaller than that
of binary-single scattering experiments. The average of $\Delta$ in
binary-binary interactions is about $0.5$.

The dimensionless rate $R_{\rm bb}$ is expressed as
\begin{equation}
 R_{\rm bb} (x_1,x_2) = 
  \frac{\sqrt{2}}{3} \frac{1}{\pi a^2 v_{\rm s,c}}
  \int v \sigma j(v) dv,
\end{equation}
where $a$ is semi-major axis of the binary with the binding energy
$x_1+x_2$, $v$ is the relative velocity between binaries, $j(v)$ is
the distribution of the relative velocity between the binaries, and
$\sigma$ is cross section of the binary-binary interactions. Assuming
that the velocity distribution of the single stars and binaries in the
core is isotropic Maxwellian, and equipartition is achieved:
$v_{\rm b,c} = (1/\sqrt{2})v_{\rm s,c}$, where $v_{\rm b,c}$ is one
dimensional velocity dispersion of the binaries, the relative
velocity dispersion between the binaries $\sqrt{2}v_{\rm b,c}$ is
$v_{\rm s,c}$, and $j(v)$ is expressed as
\begin{equation}
 j(v) = \left(\frac{2}{\pi}\right)^{\frac{1}{2}}\frac{v^2}{v_{\rm
  s,c}^3} \exp \left( - \frac{v^2}{{2v_{\rm s,c}^2}}\right).
\end{equation}
The cross section of the binary-binary interaction $\sigma$ is expressed
\begin{equation}
 \sigma = S \frac{G^2m^3}{v^2E_{\rm bin,2}},
  \label{eq:sigmabb}
\end{equation}
where $E_{\rm bin,2}$ is the binding energy of the softer binary,
expressed as $ x_2 m v_{\rm s,c}^2 / C$, and $S$ is a dimensionless
coefficient. The cross section $\sigma$ is derived, based on equation
(2.7) in Gao et al. (1991). The dimensionless coefficient $S$ depends on
the relation between the binding energies of the binaries, $x_1$ and
$x_2$, as follows:
\begin{equation}
S = \left\{
     \begin{array}{ll}
      \displaystyle 25.2 & \mbox{($x_1 \sim x_2$)} \\
      \displaystyle 15.9 & \mbox{($x_1 \gg x_2$)} \\
     \end{array}
    \right.
\end{equation}
(\cite{Gao+91}).

In summary, the dimensionless rate $R_{\rm bb}$ is expressed as
\begin{equation}
 R_{\rm bb} (x_1,x_2) = 0.479 S \left(\frac{x_1+x_2}{C}\right)
  \left[\frac{(x_1+x_2)^2}{x_2}\right],
\end{equation}
and the dimensionless hardening rate $G(x_1,x_2)$ is expressed as
\begin{equation}
 G(x_1,x_2) = 0.398 S \frac{x_1+x_2}{x_2}.
\end{equation}

We describe the forms of $h(x)$ and $H(x_1,x_2)$. When we obtain
$h(x)$, we assume that at every binary-single encounter the binary
increases its binding energy, $E_{\rm bin}$, by $0.4 E_{\rm bin}$, and
two thirds of the increment go to the kinetic energy of the single
star and the rest to the kinetic energy of the center of mass of the
binary. The increment is the average value over all the binary-single
encounters, which has been obtained by \citet{Heggie75}. The single star
will be ejected when $E_{\rm b}/m|\Phi_{\rm c}|>15/4$. The binary will
be ejected when $E_{\rm b}/m|\Phi_{\rm c}|>15$. Since the ejection
results in the mass loss of the core, the binding energy of the core
decreases, {\it i.e.} the core is heated. The amount of the heating is
$m|\Phi_{\rm c}|$ when a single star is ejected, and $2m|\Phi_{\rm c}|$
when a binary is ejected. Therefore, we can express $h(x)$ as
\begin{equation}
 h(x) = \left\{
	 \begin{array}{ll}
	  \displaystyle 1, & 
	   \mbox{if $\displaystyle x < \frac{15}{4} \frac{m|\Phi_{\rm
	   c}|}{kT}$ ;}\\
	  \displaystyle \frac{1}{3} + \frac{m|\Phi_{\rm c}|}{\bar{\Delta
	   E_{\rm bs}}}, & \mbox{if $\displaystyle \frac{15}{4}
	   \frac{m|\Phi_{\rm c}|}{kT} < x < 15 \frac{m|\Phi_{\rm
	   c}|}{kT}$;}\\
	  \displaystyle \frac{3m|\Phi_{\rm c}|}{\bar{\Delta E_{\rm
	   bs}}}, & \mbox{if $\displaystyle x > 15 \frac{m|\Phi_{\rm
	   c}|}{kT}$}. 
	 \end{array}
	\right.
\end{equation}
We set $m|\Phi_{\rm c}|=10kT$ as described in section
\ref{sec:interpretation}. Then, $h(x)$ is expressed as
\begin{equation}
 h(x) = \left\{
	 \begin{array}{ll}
	  \displaystyle 1, & \mbox{if $x < 38$ ;}\\
	  \displaystyle \frac{1}{3} + \frac{25}{x}, & \mbox{if
	   $38<x<150$;}\\
	  \displaystyle \frac{75}{x}, & \mbox{if $x>150$}.
	 \end{array}
	\right.
 \label{eq:hx}
\end{equation}

For $H(x_1,x_2)$, we assume that at every binary-binary interaction,
$0.5(E_{{\rm bin},1}+E_{{\rm bin},2})$ is liberated, either binary is
destroyed, and its $1/4$ goes to the kinetic energy of the center of
mass of the surviving binary, and its $3/8$ goes to each single star
which is a component of the destroyed binary. Then, $H(x_1,x_2)$ is
expressed as
\begin{equation}
 H(x_1,x_2) = \left\{
	       \begin{array}{ll}
		\displaystyle 1, & \mbox{if $\displaystyle (x_1+x_2) <
		 \frac{16}{3} \frac{m|\Phi_{\rm c}|}{kT}$ ;}\\
		\displaystyle \frac{1}{4} + \frac{2m|\Phi_{\rm c}|}{\bar{\Delta
		 E_{\rm bb}}}, & \mbox{if $\displaystyle \frac{16}{3}
		 \frac{m|\Phi_{\rm c}|}{kT} < (x_1+x_2) < 16
		 \frac{m|\Phi_{\rm c}|}{kT}$;}\\
		\displaystyle \frac{4m|\Phi_{\rm c}|}{\bar{\Delta E_{\rm
		 bb}}}, & \mbox{if $\displaystyle (x_1+x_2) > 16
		 \frac{m|\Phi_{\rm c}|}{kT}$},
	       \end{array}
	      \right.
\end{equation}
and substituting $m|\Phi_{\rm c}|=10kT$ we finally obtain
\begin{equation}
 H(x_1,x_2) = \left\{
	       \begin{array}{ll}
		\displaystyle 1, & \mbox{if $(x_1+x_2) < 53$ ;}\\
		\displaystyle \frac{1}{4} + \frac{40}{x_1+x_2}, &
		 \mbox{if $53<(x_1+x_2)<160$;}\\
		\displaystyle \frac{80}{(x_1+x_2)}, & \mbox{if
		 $(x_1+x_2)>160$}.
	       \end{array}
	       \right.
 \label{eq:hx1hx2}
\end{equation}

\begin{figure}
\begin{center}
\FigureFile(120mm,100mm){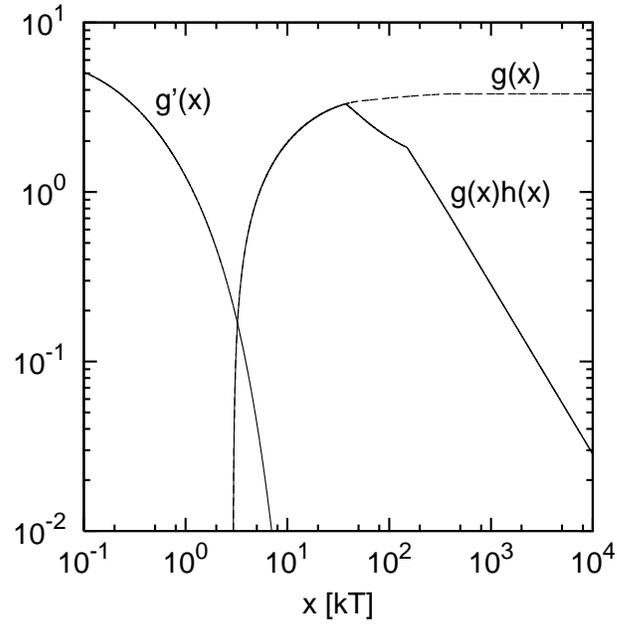}
\end{center}
\caption{Dimensionless heating rate of binary-single encounters in which
 the binaries survives ($g(x)h(x)$) and are destroyed ($g'(x)$) as a
 function of dimensionless binding energy $x$
 (solid curve). The dashed line shows the dimensionless hardening rates
 of the binary with the binding energy $x$ in a sea of single stars,
 {\it i.e.} $g(x)$.}
\label{fig:abs}
\end{figure}

\begin{figure}
\begin{center}
\FigureFile(120mm,100mm){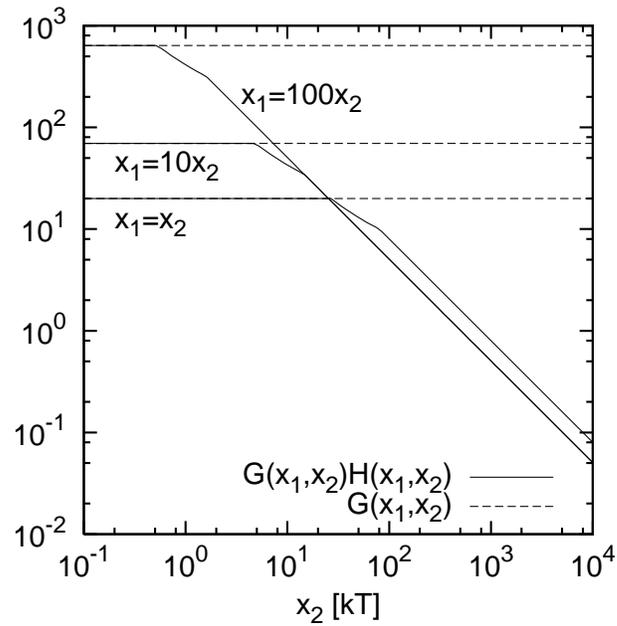}
\end{center}
\caption{Dimensionless heating rate of binary-binary encounters,
 $G(x_1,x_2)H(x_1,x_2)$, as a function of $x_2$, where the binaries have
 dimensionless binding energy $x_1$ and $x_2$, and $x_1=x_2$,
 $x_1=10x_2$, and $x_1=100x_1$ (solid curves). The dashed lines show the
 dimensionless hardening rates of the binary with the binding energy
 $x_1$ in a sea of binaries with the binding energy $x_2$,
 {\it i.e.} $G(x_1,x_2)$.}
\label{fig:abb}
\end{figure}

We show the functions $g'(x)$ and $g(x)h(x)$ in figure \ref{fig:abs},
and the function $G(x_1,x_2)H(x_1,x_2)$ in figure \ref{fig:abb}. These
functions are required for calculating the dimensionless efficiency
factors, expressed as equation (\ref{eq:Abs}) and
(\ref{eq:Abb}). Additionally, we show the function $g(x)$ in figure
\ref{eq:Abs} and the functions $G(x_1,x_2)$ in figure
\ref{fig:abb}. In figure \ref{fig:abs}, the $g(x)h(x)$ and $g(x)$ curves
overlap each other in $x<38$. In figure \ref{fig:abb}, the
$G(x_1,x_2)H(x_1,x_2)$ curves and $G(x_1,x_2)$ lines in $x_1=x_2$,
$x_1=10x_2$, and $x_1=100x_2$ overlap each other in $x_2<0.5$, $x_2<5$,
and $x_2<25$, respectively. Note that the dimensionless heating rate of
binary-binary encounters is not correct if $x_2$ is soft, {\it i.e.}
$x_2<3kT$. This is because the cross section of the binary-binary
interaction $\sigma$ is applicable only when the two binaries are hard.

\subsubsection{Core size at the halt of core contraction}
   \label{sec:th-si_cc}

We compare the core sizes of the clusters in our simulations with
those derived from equation (\ref{eq:theory}). When we use
equation (\ref{eq:theory}) to derive the ratio
$r_{\rm c}/r_{\rm h}$ at the halt of core contraction, we approximate
the distribution function of the binding energies $f(x)$ and the unit of
the binding energy $kT$ as $f(x) \simeq \delta (x-E_{\rm bin,0}/kT)$ and
$kT \simeq kT_0$, respectively. The theoretical estimate of the core
sizes is expressed as
\begin{eqnarray}
 \frac{r_{\rm c}}{r_{\rm h}} &=& \frac{0.0196\lambda}{\log_{10}(0.4N)}
  \left( \frac{v_{\rm s,c}}{v_{\rm h}}\right)^3
  \left( \frac{\gamma}{10} \right)
  (2-f_{\rm b,c})^4 \nonumber \\
  \times 
  &\Biggl\{&
   \frac{2f_{\rm b,c}(1-f_{\rm b,c})}{(2-f_{\rm
   b,c})^2}\left[g(E_{\rm bin,0}/kT)h(E_{\rm bin,0}/kT)-g'(E_{\rm
	    bin,0}/kT)\right] \nonumber \\
   &+&
   \frac{1}{2}\left(\frac{f_{\rm b,c}}{2-f_{\rm
	       b,c}}\right)^2G(E_{\rm bin,0}/kT,E_{\rm
   bin,0}/kT)H(E_{\rm bin,0}/kT,E_{\rm bin,0}/kT)
  \Biggr\}.
  \label{eq:theorydelta}
\end{eqnarray}

We justify the approximation that
$f(x) \simeq \delta (x-E_{\rm bin,0}/kT)$ and $kT \simeq kT_0$ as
follows. Figure \ref{fig:edis_tcc} shows the distributions of the
binding energies of binaries in the clusters of models $10kT_0-0.1$,
$30kT_0-0.1$, $100kT_0-0.1$, and $300kT_0-0.1$ at the time indicated in
these panels, {\it i.e.} the time when the core contractions stop. All
the distributions of the binding energies have steep peaks at the
initial binding energies, $E_{\rm bin,0}$. Even inside the core radii,
all the distributions of the binding energies may do so. The
distributions of the binding energies of binaries in the core at the
halt of the core contraction can be also regarded as the same as the
initial distributions, {\it i.e.} delta functions.

The average kinetic energies of single stars within the half-mass radii
and of binaries within the core radii are not so different from the
initial time to the time at the halt of the core contraction, as seen in
figure \ref{fig:ekinsin} and \ref{fig:ekinbin}, respectively. We can
regard $1kT=1kT_0$.

\begin{figure}
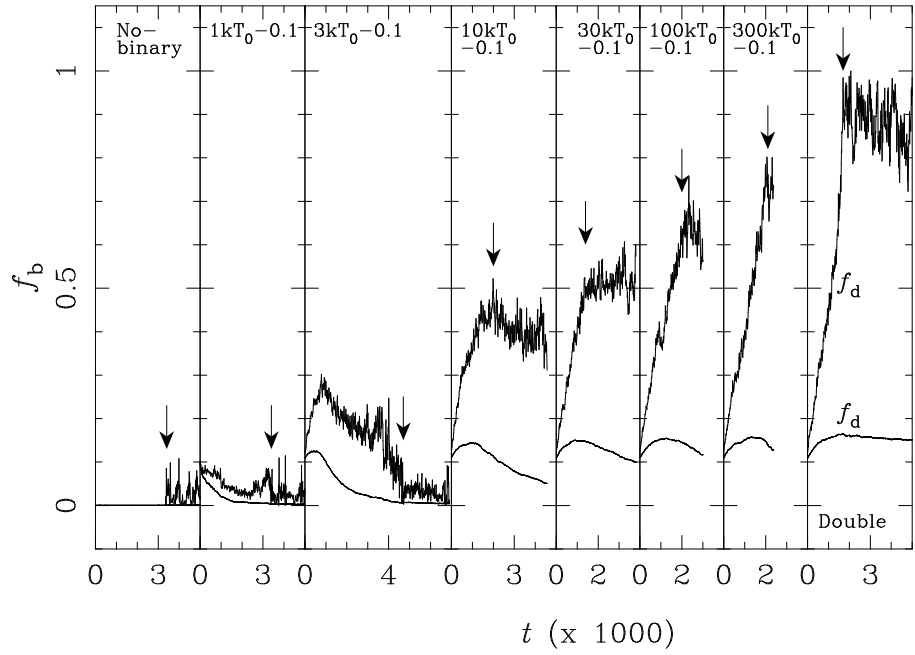

\begin{center}
\FigureFile(120mm,100mm){figure8.ps}
\end{center}
\caption{Time evolution of the mass fraction, $f_{\rm b}$, of the
 binaries inside the core radii (the upper curves) and  half-mass radii (the
 lower curves) in the $f_{\rm b,0}=0.1$ models and model No-binary. For
 model Double, the mass fraction, $f_{\rm d}$, of the double mass stars
 are shown. The arrows show the time when the core collapse stops.}
\label{fig:binfrac}
\end{figure}

\begin{figure}
\begin{center}
\FigureFile(120mm,125mm){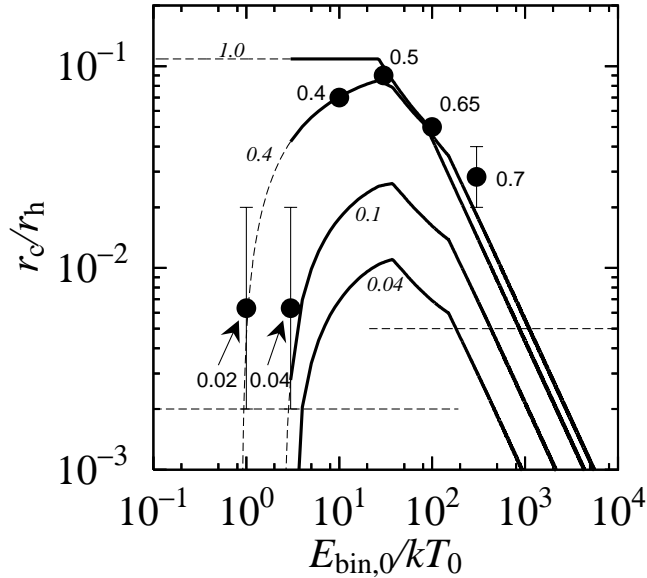}
\end{center}
 \caption{Ratio of core radii to half-mass radii at the halts of core
 collapse of the clusters whose primordial binaries have equal binding
 energy, $E_{\rm bin}$, and whose cores contain the mass fraction of the
 primordial binaries in the core, $f_{\rm b,c}$. The dots show the
 ratio of the core radii to the half-mass radii at the halt of core
 collapse in our simulation. The numbers beside the dots are $f_{\rm
 b,c}$ at that time. The four curves draw equation (\ref{eq:theory}) when
 $f_{\rm b,c}=0.04$, $0.1$, $0.4$, and $1.0$. The values of $f_{\rm
 b,c}$ are beside the curves in italic format. The dashed lines, $r_{\rm
 c}/r_{\rm h}=0.002$, and $0.005$, show the ratio of core radii to
 half-mass radii at the halts of core collapse in model No-binary and
 Double, respectively.}
\label{fig:rc-kt}
\end{figure}

In figure \ref{fig:rc-kt}, the big black dots show $r_{\rm c}/r_{\rm h}$
at the halts of core collapse of the clusters obtained in our
simulations, as a function of the initial binding energy in the unit of
$kT$, $E_{\rm bin,0}/kT$. The numbers beside the dots show the mass
fraction of the binaries in the core, $f_{\rm b,c}$, at the time of the
halt, obtained from figure \ref{fig:binfrac}. The error bars indicate
the amplitude of gravothermal oscillations. The dots are the geometric
means of the maximum and minimum $r_{\rm c}/r_{\rm h}$ in the
gravothermal oscillations. Solid curves show equation
(\ref{eq:theorydelta}) when $f_{\rm b,c}=0.04,0.1,0.4,1$, which is shown
by the numbers in italic format beside the curves. Here, we adopt
$v_{\rm s,c}/v_{\rm h}=\sqrt{2}$, $\gamma=10$, ant the free parameter
$\lambda=0.45$. In the binding energy with less than
$E_{\rm bin} \lesssim E_{\rm kin,ave,s} \sim 2.5kT_0$, the curves are
not reliable, and should fall down to zero. Figure \ref{fig:rc-kt} shows
that the pairs of the ratios $r_{\rm c}/r_{\rm h}$ and the binary
fraction in the core $f_{\rm b,c}$ at the halts of core contraction for
models $10kT_0-0.1$, $30kT_0-0.1$, $100kT_0-0.1$, and $300kT_0-0.1$ in
our simulations are included in those expressed by equation
(\ref{eq:theorydelta}).

The pairs of the ratios $r_{\rm c}/r_{\rm h}$ and the binary fraction in
the core $f_{\rm b,c}$ at the halts of core contraction for models
$1kT_0-0.1$ and $3kT_0-0.1$ are not included in those expressed by
equation (\ref{eq:theorydelta}), and the ratios
$r_{\rm c}/r_{\rm h}$ are much larger than equation
(\ref{eq:theorydelta}) despite of their small binary fraction in the
cores. This is because the core collapses of the clusters stop due to
the energy that goes to the cluster from the three-body binaries
composed of the single mass stars. The three-body binaries appear after
the deep core collapse as discussed in section \ref{sec:mainresults}.

We compare the pairs of the ratios $r_{\rm c}/r_{\rm h}$ and the binary
fraction in the core $f_{\rm b,c}$ at the halts of core contraction
predicted by equation (\ref{eq:theorydelta}) with those
for model Double as the hard limit of the primordial binaries. Equation
(\ref{eq:theorydelta}) should not include the simulation results of
model Double. Equation (\ref{eq:theorydelta}) indicates that the ratio
$r_{\rm c}/r_{\rm h}$ is zero, regardless of the mass fraction of the
primordial binaries in the core, $f_{\rm b,c}$ (see solid curves in
figure \ref{fig:rc-kt}). On the other hand, $r_{\rm c}/r_{\rm h}=0.005$
and $f_{\rm b,c} \sim 1$ at the halt of the core contraction in model
Double.

The reason for such disagreement is as follows. In the theoretical
estimate which derives equation (\ref{eq:theorydelta}), only primordial
binaries are considered as energy sources for the cluster. However, in
model Double, another energy source appears. The energy source is
binaries consisting of two double mass stars. The binaries are formed
through the encounters of three single stars with double mass. The thick
curve in model Double in figure \ref{fig:energetics} shows that such
binaries generate energy.

We expect that, in cluster models with harder primordial binaries than
those we treat, similar energy sources to binaries consisting of two
double mass stars in model Double appear. The energy sources are
hierarchical quadruple systems in which two binaries orbit around each
other. The hierarchical quadruple systems may generate energy by
shrinking orbits of the two binaries through interactions with the
surrounding stars.

We estimate the critical binding energy of the primordial binaries in
which the hierarchical quadruple systems are formed, generate energy,
and stop the contraction of the cluster core. We expect that the
hierarchical quadruple systems are formed when
$r_{\rm c}/r_{\rm h} \sim 0.005$. This is because the binaries
consisting of two double mass stars are formed at such core size in
model Double. When $E_{\rm bin} \sim 1000kT_0$, the core contraction
stops at $r_{\rm c}/r_{\rm h} \sim 0.005$ as seen in solid lines of
$f_{\rm b,c}=0.4$ and $1.0$ in figure \ref{fig:rc-kt}. It is not
necessary to consider low mass fraction of the primordial binaries in
the core, such as $f_{\rm b,c}<0.1$, since the primordial binaries
become more centrally-concentrated as they becomes harder (see figure
\ref{fig:binfrac}). Therefore, the critical binding energy is
$\sim 1000kT_0$.

\subsection{Initial mass fraction}

\begin{figure}
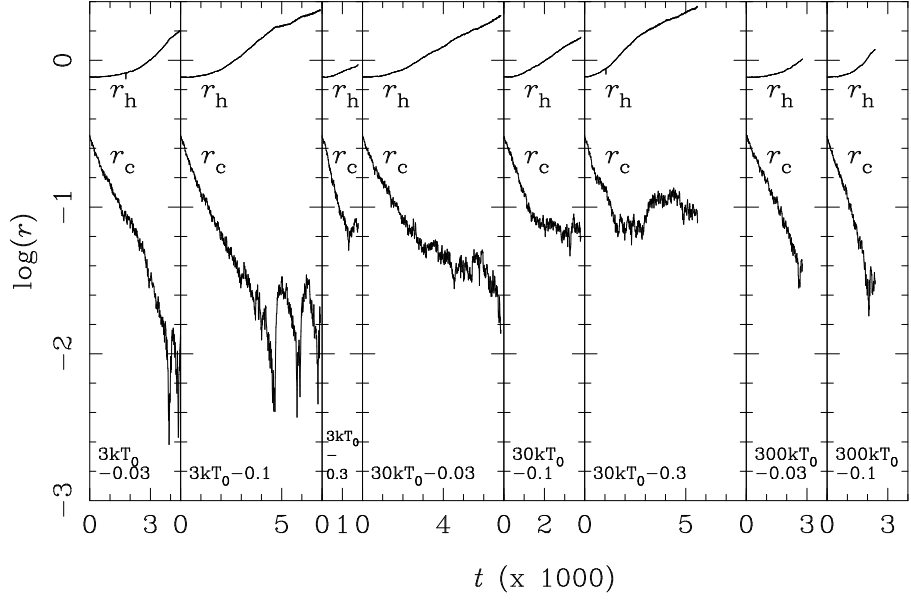

\begin{center}
\FigureFile(120mm,80mm){figure16.ps}
\end{center}
\caption{Time evolution of the core, $r_{\rm c}$, and half-mass radii,
 $r_{\rm h}$, of the clusters with $f_{{\rm b},0}=0.03$, $0.1$, and $0.3$
 primordial binaries, each of which has the binding energy $E_{{\rm
 bin},0}=3kT_0$, $10kT_0$, and $300kT_0$. The way of calculation of the core
 radii is the same as that in figure \ref{fig:rc_all-t}.}
\label{fig:rcrh_3e012kt}
\end{figure}

In this section, we discuss the dependence of the core evolution on the
initial mass fraction of the primordial binaries, $f_{\rm b,0}$. Figure
\ref{fig:rcrh_3e012kt} shows the time evolution of the core radii,
$r_{\rm c}$, and half-mass radii, $r_{\rm h}$, of the clusters with
$f_{\rm b,0}=0.03$, $0.1$, and $0.3$ primordial binaries, each of which
has the binding energy $E_{\rm bin,0}=3kT_0$, $30kT_0$, and $300kT_0$. In
models $3kT_0-0.03$, and $3kT_0-0.1$, deep core collapse occurs, and in
model $3kT_0-0.3$, core collapse stops halfway. In all
$E_{\rm bin,0}=30kT_0$ models, core collapse stops halfway. In all
$E_{\rm bin,0}=300kT_0$ models, deep core collapse occurs.

Figure \ref{fig:binfrac_3e012kt} shows the time evolution of the mass
fraction of the binaries inside the core and half-mass radii, $f_{\rm
b}$, of the clusters in the $f_{\rm b,0}=0.03$, $0.1$, and $0.3$ models,
each of which has the binding energy $E_{\rm
bin,0}=3kT_0$, $30kT_0$, and $300kT_0$. In all of them, mass segregation
occurs initially. However, the mass fraction of the binaries in the core
decrease halfway in models $3kT_0-0.03$, $3kT_0-0.1$, and $30kT_0-0.03$.
In models $30kT_0-0.03$, and $300kT_0-0.03$, the turning
points correspond to the time when core collapse stops. In contrast, in
models $3kT_0-0.03$, and $3kT_0-0.1$, the mass fractions of the binaries
in the core decrease long before the core collapse.

In contrast to models $3kT_0-0.03$, and $3kT_0-0.1$, the core collapse
of the cluster in model $3kT_0-0.3$ stops halfway. Since the cluster has
many primordial binaries, the energy from the primordial binaries to the
cluster is large enough to stop the core collapse.

\begin{figure}
\begin{center}
\FigureFile(120mm,125mm){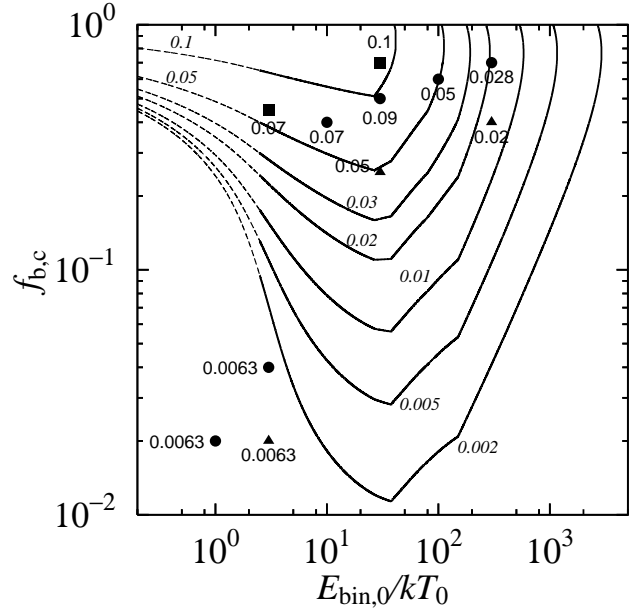}
\end{center}
 \caption{Contours of the ratio of core radii to half-mass radii at the
 halt of core contraction in clusters whose mass fraction of primordial
 binaries in the core is $f_{\rm b,c}$, and distribution function of the
 primordial binaries is $\delta (x-E_{\rm bin})$, which are obtained
 from equation (\ref{eq:theorydelta}). The numbers in italic  show the
 ratio
 of the core radii to the half-mass radii. The dashed  curves are
 possibly not correct, since the heating rate through  binary-binary
 interactions in hard binaries is extended to the soft range. The dots
 show simulation results. The vertical axis shows $f_{\rm b,c}$ at the
 halt of core contraction, the horizontal axis shows $E_{\rm bin}$ at
 the initial time, and the numbers beside the dots is the ratio of core
 radii to the half-mass radii at the halt of the core contraction. The
 shapes of the dots show the initial mass fraction of the primordial
 binaries in the clusters, $f_{{\rm b},0}$. The triangles, circles, and
 squares show the models $f_{{\rm  b},0}=0.03$, $0.1$, and $0.3$,
 respectively.}
\label{fig:phi-kt}
\end{figure}

We compare the ratios of the core radii to the half-mass radii in our
simulation results with the theoretically estimated ratio in equation
(\ref{eq:theory}). The dots in figure \ref{fig:phi-kt} show the mass
fraction of the binaries in the core at the halt of core collapse,
$f_{\rm b,c}$, of the clusters whose primordial
binaries have the initial binding energy, $E_{\rm bin,0}$. The
numbers beside the dots show the ratios of the core radii
to the half-mass radii at the halt of core collapse. When gravothermal
oscillations occur, the geometric means are shown. The triangles,
circles, and squares show the models $f_{\rm b,0}=0.03$, $0.1$, and
$0.3$, respectively.

The curves in figure \ref{fig:phi-kt} show the ratios of the core radii
to the half-mass radii, $r_{\rm c}/r_{\rm h}$, of clusters whose
primordial binaries have the initial binding energy, $E_{\rm bin,0}$,
and whose mass fraction of the binaries in the core is $f_{\rm b,c}$ at
the halt of the core collapse. The numbers in italic format beside the
curves indicate the values of $r_{\rm c}/r_{\rm h}$. If
$E_{\rm bin,0}\le2.5kT_0$, the curves are not reliable. This is the same
reason as in figure \ref{fig:rc-kt}.

In the clusters above the curve of $r_{\rm c}/r_{\rm h}=0.002$, the core
collapse stops halfway, and in the clusters below the curve of $r_{\rm
c}/r_{\rm h}=0.002$, the clusters experience deep core collapse. In the
models $E_{\rm bin,0}=3kT_0$, model $3kT_0-0.03$ and $3kT_0-0.1$ is
below the curve, and model $3kT_0-0.3$ is above the curve. This is in
good agreement with our simulation results. The ratios $r_{\rm c}/r_{\rm
h}$ in models $1kT_0-0.1$, $3kT_0-0.03$, and $3kT_0-0.1$ disagree with
theoretical curves, since the core collapse stops due to energy heating
core generated by the three-body binaries.

\subsection{High-velocity escapers}

We investigate escapers of each cluster in the $f_{\rm b,0}=0.1$
models. Figure \ref{fig:singleandbinary} show the number of
single escapers ($N_{\rm esc,sin}$: solid lines) and binary escapers
($N_{\rm esc,bin}$: dotted lines) in each logarithmic bin of velocities
of the single escapers ($v_{\rm sin}$) and the binary escapers ($v_{\rm
bin}$) for all $f_{\rm b,0}=0.1$ models. 
The velocities of these escapers are measured at the moment when they
satisfy the conditions of the escapers, as shown in section
\ref{sec:mainresults}.
Note that in model Double, the single and binary escapers correspond to
the escapers of the single and double mass stars, respectively. In the
single escapers for all models except model Double, two peaks are
present, although the higher peaks are small in models No-binary,
$1kT_0-0.1$, and $3kT_0-0.1$. The population of the lower velocity
escapers is driven by two-body relaxation, and that of the higher
velocity escapers is ejected from the clusters through binary-single and
binary-binary encounters. In the binary escapers, the escape velocities
are similar to high-velocity population of the single escapers. They are
also ejected from the clusters through binary-single and binary-binary
encounters.

\begin{figure}
\begin{center}
\FigureFile(100mm,75mm){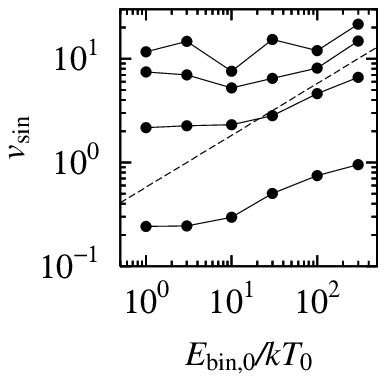}
\end{center}
 \caption{The largest, and the $0.01N_{\rm esc,sin}$-th, $0.1N_{\rm
 esc,sin}$-th, and $0.5N_{\rm esc,sin}$-th largest velocities of single
 escapers from top to bottom as a function of initial binding energy of
 primordial binaries in the $f_{{\rm b},0}=0.1$ clusters. The dashed
 line shows circular velocity of a binary as a function of its binding
 energy.}
 \label{fig:vmeasures_single}
\end{figure}

\begin{figure}
\begin{center}
\FigureFile(100mm,75mm){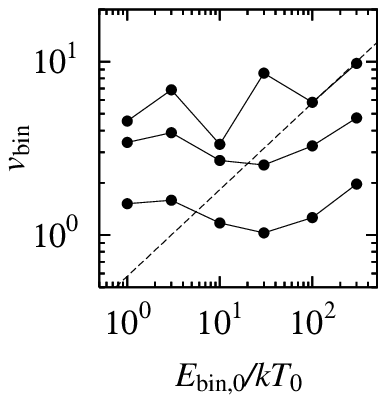}
\end{center}
 \caption{The largest, and the $0.1N_{\rm esc,bin}$-th and
 $0.5N_{\rm esc,bin}$-th largest velocities of binary escapers from top
 to bottom as a function of initial binding energy of primordial
 binaries in the $f_{{\rm b},0}=0.1$ clusters. The dashed line shows
 circular velocity of a binary as a function of its binding energy.}
 \label{fig:vmeasures_binary}
\end{figure}

Figure \ref{fig:vmeasures_single} shows the largest and the
top $1$ per cent ($0.01 N_{\rm esc,sin}$-th largest), top $10$ per cent
($0.1 N_{\rm esc,sin}$-th largest), and top $50$ per cent
($0.5 N_{\rm esc,sin}$-th largest) velocities of single escapers from
top to bottom in as a function of the initial binding energy of the
primordial binaries, $E_{\rm bin,0}$, in the models
$f_{{\rm b},0}=0.1$. Figure \ref{fig:vmeasures_binary} shows those of
the binary escapers, although the top $1$ per cent
($0.01 N_{\rm esc,bin}$-th largest)velocity is omitted because of the
small number of the binary escapers. The dashed lines in both figures
show circular velocities of the binaries as a function of the initial
binding energy, $E_{\rm bin,0}$.

Consider a globular cluster whose virial radius is $10$ pc, and whose
mass is $10^6$ solar mass. Then, one velocity unit is $24 kms^{-1}$. When we
apply our simulation results for the cluster, the highest velocity of
the escapers is $500 km s^{-1}$, which is the single escaper in model
$300kT_0-0.1$. In order to form hyper-velocity stars, which orbit in our
Galaxy at speeds of $500-1000 kms^{-1}$ (\cite{Hirsch+05};
\cite{Brown+05}; \cite{Edelmann+05}; \cite{Heber+08}), the globular
clusters have to contain a large fraction of hard binaries with
$\sim 300kT_0$ at the initial time. The presence of many hard binaries
with $\sim 300kT_0$ at the initial time is possible, since $300kT_0$
binaries are contact binary when the binary components are main-sequence
stars.

\begin{figure}
\begin{center}
\FigureFile(120mm,75mm){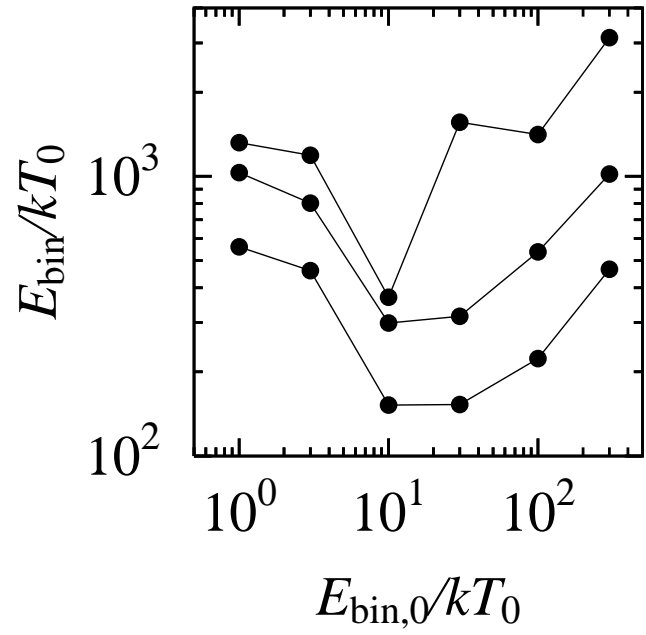}
\end{center}
 \caption{The largest, and the $0.1N_{\rm esc,bin}$-th and
 $0.5N_{\rm esc,bin}$-th largest binding energies of binary escapers
 from top to bottom as a function of initial binding energy of
 primordial binaries in the $f_{{\rm b},0}=0.1$ clusters.}
 \label{fig:ebinmeasures}
\end{figure}

Figure \ref{fig:escbinebin} shows the number of binary escapers
($N_{\rm esc,bin}$) in each logarithmic bin of binding energies of the
binary escapers ($E_{\rm bin}$) for model No-binary and all
$f_{\rm b,0}=0.1$ models. Although the total numbers of the binary
escapers are different among model No-binary and all $f_{\rm b,0}=0.1$
models, the binding energies of most binary escapers ranges from
$100kT_0$ to $1000kT_0$. Figure \ref{fig:ebinmeasures} shows the
largest, and the top $10$ per cent ($0.1 N_{\rm esc,bin}$-th largest)
and top $50$ per cent ($0.5 N_{\rm esc,bin}$-th largest) binding
energies of the binary escapers from top to bottom as a function of the
initial binding energies, $E_{\rm bin,0}$ for all $f_{{\rm b},0}=0.1$
models.

Additionally, we list the triple escapers in table \ref{tab:triple}. The
second column is the velocity of the center of mass of the triple
escapers, $v_{\rm tri}$. The third and fourth columns are, respectively,
the binding energy ($E_{\rm bin,in}$ and $E_{\rm bin,out}$) of the inner
and outer binaries in unit of $kT_0$.

\section{Summary}
\label{sec:summary}

We study systematically the dependence of cluster evolution on the
binding energy of primordial binaries. By means of GORILLA, we simulate
the core evolution of the clusters, each of which contains primordial
binaries with equal binding energy.

When the initial mass fraction of the primordial binaries is fixed to
$0.1$, we find that the dynamical evolutions of the clusters are divided
into three ranges according to hardness of the primordial binaries as
follows.
\begin{enumerate}
 \item
   In soft range ($<3kT_0$), the clusters experience core collapse
   in similar way to those without primordial binaries. The ratios of
   core radii to half-mass radii at the halt of the core collapse are
   about $0.006$. The
   primordial binaries do not heat the clusters. This is because the
   primordial binaries are destroyed through encounters with single
   stars, and do not generate energy.
 \item
   In intermediate hard range ($10kT_0-100kT_0$), the core collapses
   in the clusters halt halfway. The ratios of core radii to half-mass
   radii at the halt of the core collapse are $0.05-0.1$. The
   primordial binaries release energy, and the energy heats the
   clusters.
 \item 
   In super hard range ($>300kT_0$), the clusters experience core
   collapse, and the ratios of core radii to half-mass radii at the
   halt of the core collapse is
   about $0.02$. The primordial binaries do not so much heat the
   clusters. Although the primordial binaries release energy
   through encounters, the energy is so large that binaries and
   single stars involved with the encounters are ejected from the
   clusters.
\end{enumerate}

The dependences of the boundaries between the soft and intermediate hard
ranges and between the intermediate and super hard ranges on the initial
mass fraction of the primordial binaries are as follows.
\begin{enumerate}
 \item The boundary between the soft and intermediate hard ranges
       depends on the initial mass fraction of the primordial
       binaries. When the mass fraction of the primordial binaries is
       $0.3$, the core contraction in the cluster with $3kT_0$
       primordial binaries halts at large ratio of core radius to
       half-mass radius $\sim 0.07$, and
       the intermediate hard range includes $3kT_0$.
 \item The boundary between the intermediate and super hard ranges is
       not changed, when the initial mass fraction of the primordial
       binaries ranges from $0.03$ to $0.3$.
\end{enumerate}

We compared the pairs of the ratios of core radii to half-mass radii and
the core mass fraction of the binaries at the halt of the core
contraction in our simulations with those of theoretical estimates. We
found a good agreement between $N$-body simulations and the theoretical
values.

\section*{Acknowledgement}

We are grateful to Junichiro Makino for helpful advice. A. Tanikawa is
financially supported by Research Fellowships of the Japan Society for
the Promotion of Science for Young Scientist. This research was
supported by the Research for the Future Program of Japan Society for
the Promotion of Science (JSPS-RFTF97P01102), the Grants-in-Aid by the
Japan Society for the Promotion of Science (14740127) and by the
Ministry of Education, Science, Sports, and Culture of Japan
(16684002). Numerical computations were in part carried out on GRAPE
system at Center for Computational Astrophysics, CfCA, of National
Astronomical Observatory of Japan.

\newpage

\begin{table}
 \caption{Initial models.}
 \begin{center}
  \begin{tabular}{crlrlr}
   \hline
   \hline
   Model name & $E_{\rm bin,0}$ & $f_{\rm b,0}$ & $N_{\rm b,0}$ &
   $f_{\rm d,0}$ & $N_{\rm d,0}$ \\
   \hline
   $1kT_0-0.1$   & $1kT_0$    & 0.1 & 819 & 0 & 0 \\
   $3kT_0-0.1$   & $3kT_0$    & 0.1 & 819 & 0 & 0 \\
   $10kT_0-0.1$  & $10kT_0$   & 0.1 & 819 & 0 & 0 \\
   $30kT_0-0.1$  & $30kT_0$   & 0.1 & 819 & 0 & 0 \\
   $100kT_0-0.1$ & $100kT_0$  & 0.1 & 819 & 0 & 0 \\
   $300kT_0-0.1$ & $300kT_0$  & 0.1 & 819 & 0 & 0 \\
   \hline
   $3kT_0-0.03$  & $3kT_0$   & 0.03 & 246 & 0 & 0 \\
   $30kT_0-0.03$ & $30kT_0$  & 0.03 & 246 & 0 & 0 \\
   $300kT_0-0.03$& $300kT_0$ & 0.03 & 246 & 0 & 0 \\
   \hline
   $3kT_0-0.3$   & $3kT_0$   & 0.3  & 2458 & 0 & 0 \\
   $30kT_0-0.3$  & $30kT_0$  & 0.3  & 2458 & 0 & 0 \\
   \hline
   No-binary    & $-$     & 0   & 0   & 0 & 0 \\
   Double       & $-$     & 0   & 0   & 0.1 & 819 \\
   \hline
  \end{tabular}
 \end{center}
\label{tab:initialmodel}
\end{table}

\begin{table}
 \caption{Accuracy, and apocentric and pericentric parameters.}
 \begin{center}
  \begin{tabular}{ccccc}
   \hline
   \hline
   Model name  & $\eta$ & $\eta_s$ & $\alpha$ & $\beta$ \\
   \hline
   $1kT_0-0.1$   & $0.01$ & $0.0025$ & $5$      & $10$    \\
   $3kT_0-0.1$   & $0.01$ & $0.0025$ & $5$      & $10$    \\
   $10kT_0-0.1$  & $0.01$ & $0.0025$ & $5$      & $10$    \\
   $30kT_0-0.1$  & $0.01$ & $0.0025$ & $5$      & $10$    \\
   $100kT_0-0.1$ & $0.01$ & $0.0025$ & $5$      & $50$    \\
   $300kT_0-0.1$ & $0.01$ & $0.0025$ & $8$      & $50$    \\
   \hline
   $3kT_0-0.03$  & $0.01$ & $0.0025$ & $5$      & $10$    \\
   $30kT_0-0.03$ & $0.01$ & $0.0025$ & $5$      & $10$    \\
   $300kT_0-0.03$& $0.01$ & $0.0025$ & $5$      & $50$    \\
   \hline
   $3kT_0-0.3$   & $0.01$ & $0.0025$ & $5$      & $10$    \\
   $30kT_0-0.3$  & $0.01$ & $0.0025$ & $5$      & $50$    \\
   \hline
   No-binary   & $0.01$ & $0.0025$ & $5$      & $10$    \\
   Double      & $0.01$ & $0.0025$ & $5$      & $10$    \\
   \hline
  \end{tabular}
 \end{center}
\label{tab:parameters}
\end{table}

\begin{table}
 \caption{The list of triple escapers.}
 \begin{center}
  \begin{tabular}{clcc}
   \hline
   \hline
   Model name & $v_{\rm tri}[$standard units$]$ &
   $E_{\rm bin,in}[kT_0]$ & $E_{\rm bin,out}[kT_0]$ \\
   \hline
   $30kT_0-0.1$ & $4.7$ & $8.5\times 10^2$ & $3.0 \times 10^1$ \\
   \hline
   $100kT_0-0.1$ & $0.44$ & $2.8 \times 10^2$ & $9.2 \times 10^0$ \\
               & $1.2$  & $5.7 \times 10^2$ & $7.4 \times 10^0$ \\
               & $0.39$ & $4.7 \times 10^2$ & $2.6 \times 10^0$ \\
               & $1.7$  & $6.2 \times 10^2$ & $1.5 \times 10^0$ \\
   \hline
   $300kT_0-0.1$ & $2.0$  & $9.4 \times 10^2$ & $3.7 \times 10^1$ \\
   \hline
  \end{tabular}
 \end{center}
 \label{tab:triple}
\end{table}

\begin{figure}
\begin{center}
\FigureFile(120mm,100mm){figure10.ps}
\end{center}
\caption{Time evolution of the mean kinetic energy of the binaries,
 $E_{\rm kin,ave,b}$, inside the core radii in the $f_{{\rm b},0}=0.1$
 models. The arrows show the time when the core collapse stops.}
\label{fig:ekinbin}
\end{figure}

\begin{figure}
 \begin{center}
  \FigureFile(100mm,60mm){figure14.ps}
 \end{center}
 \caption{Distributions of the binding energies of binaries in the whole
 clusters at the time indicated in each panel, which is the time when
 the core contractions stop.}
 \label{fig:edis_tcc}
\end{figure}

\begin{figure}
\begin{center}
\FigureFile(120mm,80mm){figure17.ps}
\end{center}
\caption{Time evolution of the mass fraction of the binaries inside the
 core and half-mass radii of the clusters, $f_{\rm b}$, with $f_{{\rm
 b},0}=0.03$, $0.1$, and $0.3$ primordial binaries, each of which has the
 binding energy $E_{{\rm bin},0}=3kT_0$, $10kT_0$, and $300kT_0$.}
\label{fig:binfrac_3e012kt}
\end{figure}

\begin{figure}
\begin{center}
\FigureFile(120mm,100mm){figure19.ps}
\end{center}
 \caption{The distribution of the velocities of single and binary
 escapers (solid and dotted lines, respectively) in the models $f_{{\rm
 b},0}=0.1$. The velocity is in $N$-body  standard units.}
 \label{fig:singleandbinary}
\end{figure}

\begin{figure}
\begin{center}
\FigureFile(120mm,100mm){figure22.ps}
\end{center}
 \caption{The distribution of the binding energy of binary
 escapers in the models $f_{{\rm b},0}=0.1$. The binding energy is in
 the unit of $kT_0$.}
 \label{fig:escbinebin}
\end{figure}

\end{document}